\documentclass[prd,nofootinbib,superscriptaddress,preprint]{revtex4}

\usepackage[T1]{fontenc}
\usepackage[utf8]{inputenc} 
\usepackage{amsmath,amssymb}
\usepackage{epsfig}
\usepackage{graphicx,array}
\usepackage[usenames,dvipsnames]{color}
\usepackage{slashed}
\usepackage{comment}
\usepackage{booktabs}
\usepackage{enumitem}

\usepackage[colorlinks,citecolor=blue]{hyperref}
\usepackage{pdfpages}
\usepackage{color}
\usepackage{subfigure}

\usepackage{multirow}
\usepackage{tikz}
\definecolor{le43}{HTML}{003F88}
\definecolor{le45}{HTML}{2CA02C}
\definecolor{le47}{HTML}{C1121F}
\definecolor{me43}{HTML}{FF7F0E}
\definecolor{me45}{HTML}{00B4D8}
\definecolor{me47}{HTML}{C77DFF}
\usepackage{xcolor}
\definecolor{myblue}{HTML}{1F77B4}
\definecolor{myorange}{HTML}{FF7F0E}
\definecolor{mygreen}{HTML}{2CA02C}
\definecolor{myred}{HTML}{D62728}
\definecolor{mypurple}{HTML}{9467BD}

\newcommand{\s}{\mbox{\tiny$S$}}

\begin{document}
\title{DUNE Beam-Energy Dependence of Sensitivity to the Violation of the Equivalence Principle}
\author{Monjowara Khatun}
\email[Email Address: ]{monjowarak@gmail.com}
\affiliation{Department of Physics, Bhattadev University, Bajali, Pathsala, 781325, India}

\author{Debajyoti Dutta$^{1}$}
\email[Email Address: ]{phy.debajyoti@bhattadevuniversity.ac.in}


\begin{abstract}
 In this work, we investigate the dependence of DUNE's sensitivity to possible violations of the equivalence principle (VEP) on the beam-energy configuration, comparing the Low-Energy (LE) and Medium-Energy (ME) beams. We find that the LE configuration generally provides stronger constraints on the VEP parameters than the ME configuration. For $\theta_{23}=43^\circ$, the LE configuration constrains $\Phi\Delta\gamma_{21}$ and $\Phi\Delta\gamma_{31}$ to $8.47\times10^{-24}$ and $8.53\times10^{-24}$, respectively, at the $3\sigma$ CL. The appearance channels shows stronger phase-dependency  degeneracies while the disappearance channels exhibits smoother exclusion sensitivity. This work further confirms the advantage of the LE configuration over the ME, in constraining the correlated VEP parameter space. For the assumed true value of $\theta_{23}=43^\circ$, The LE configuration restricts the VEP parameter space to $\mathit{\Phi}\Delta\gamma_{21}\in[-8.47,\,8.47]\times10^{-24}$ and $\mathit{\Phi}\Delta\gamma_{31}\in[-8.53,\,8.53]\times10^{-24}$ at $3\sigma$.We also find that neutrino-mode and antineutrino-mode running can independently probe VEP effects at DUNE, with the relative sensitivity depending on the beam-energy configuration. These results highlight the importance of beam-energy optimization and channel-wise analysis in exploring possible violations of the equivalence principle at DUNE.

\end{abstract}

\pacs{14.60.St,14.60.Pq,14.60.Lm,13.15.+g}
\keywords{Equivalence Principle, DUNE}
\maketitle

\section{Introduction}

The Weak Equivalence Principle (WEP) is one of the founding stones of modern physics \cite{Weinberg1972,Misner1973,  Will1993, Carroll2004}. In classical mechanics, WEP meant to be any of the following two statements:
\begin{enumerate}
\item Inertial mass $m_{i}$ is proportional to gravitational mass $m_{g}$. [In appropriate units; $m_{i}$=$m_{g}$]
\item Dynamics of a test particle in a gravitational field is independent of its intrinsic properties, e.g mass, charge, etc.; i.e all test particles fall with equal acceleration in gravity; also known as Universality of the free fall (UFF).
\end{enumerate}
But the two statements above are logically independent, and they happen to imply each other in the Newtonian framework. For our work here, we will refer first to the Newtonian Equivalence Principle (NEP) and second to WEP, following the notation from \cite{DiCasola2015}. WEP imply NEP, but the reverse need not be true. In Non-Relativistic Quantum Mechanics, if we assume NEP, the Schr$\ddot{o}$dinger equation remains dependent on the mass of the particle, but when we solve the Schr$\ddot{o}$dinger equation in the accelerated frame, it is satisfied by the wave function of a free particle (in the accelerated frame) modulo a phase factor which is a function of the mass parameter; i.e. effects of the constant gravitational field can be removed by going into an appropriate accelerated frame just like in classical mechanics \cite{Herdegen:2001tt}.
Dynamics is independent of the particle properties in Non-Relativistic Quantum Mechanics too; WEP holds. It is important to check it as we derive neutrino oscillation probabilities from the Schr$\ddot{o}$dinger equation \cite{Giunti2007}. Einstein generalised WEP to Einstein’s Equivalence Principle (EEP), which forms the foundation of all metric theories of gravity; along with WEP, validity also requires two additional conditions: that the outcome of a non-gravitational experiment is independent of the velocity of the freely falling apparatus and its position in the spacetime \cite{Williams:1995nq}. Many Beyond Standard Model (BSM) theories predict the violation of the equivalence principle. Experimental tests for Violation of the Weak Equivalence Principle (VWEP) are important, as any experimental hints about violation or no violation can help us to constrain the possible sample space of the BSM theories and might help to find out the future directions for a more fundamental theory than the Standard Model.

The idea that we can test VWEP using Neutrino Oscillation was put forth in the late 90's by Gesperini \cite{Gasperini:1988zf,Gasperini:1989rt}. Initially, when the theory of Neutrino Oscillations was not well established, it was thought of as an alternate mechanism for neutrino oscillations. VWEP was used as a possible explanation of the Solar Neutrino oscillations \cite{Bahcall:1994zw, Halprin:1991gs, Halprin:1995vg, Gago:1999hi, Mansour:1998nb}. Combined data of SNO and SK-solar data using VWEP is studied into \cite{Raychaudhuri:2001gy}. As more data accumulated over the years, it was found that VWEP alone can not explain the neutrino oscillation experiments \cite{Mann:1995nw}. Over the span of the last two decades, the leading theory of Neutrino Oscillations is well established. Our current understanding is that neutrino oscillations occur due to a non-zero, non-diagonal mixing matrix between mass and flavour eigenstates and non-zero mass differences between different mass eigenstates, i.e not all eigenvalues of the mass matrix are degenerate. If VWEP is there, it will give different anatomical signatures in the oscillation probabilities and hence in the event rates, and we 
can probe it in Neutrino experiments as a sub-leading effect. 

There exist many string-inspired models \cite{Damour1994, Damour1994a} in which one finds violations of the universality of free fall (UFF) at the level of $\eta \equiv \frac{\Delta a}{a} \sim 10^{-18}$. There are many ongoing efforts in probing these violations. The universality of free fall has been directly verified for composite bodies up to $10^{-13}$ \cite{ Schlamminger:2007ht, Williams:1995nq}. There is also test of UFF done with quantum particle; $\eta_{\rm Rb, K} \simeq (0.3 \pm 5.4) \times 10^{-7}$ \cite{Schlippert:2014xla}. There have also been studies testing UFF for different spin particles \cite{Duan2016}.

In the neutrino sector, there have been numerous studies to probe VEP. First, there are those which try to probe VEP using neutrino oscillations. In \cite{Gonzalez-Garcia:2004pka} for $\mu$-$\tau$ sector $\mathit{\Phi} \Delta\gamma_{32}< 4 \times 10^{-25}$ is reported by studying K2K and SK-Atmospheric data with $3 \s$ C.L. Analysis is being done for a 2-$\nu$ scenario. In \cite{Battistoni2005} $\mathit{\Phi} \Delta\gamma_{32}< 1.5 \times 10^{-25}$ using MACRO atmospheric data. Recently, It has been reported that using Solar, reactor data we can limit $\mathit{\Phi} \Delta\gamma < 2 \times 10^{-19}$ at 3$\s$ for $\Delta \gamma$ contributing to $\nu_e$ and $\bar{\nu_{e}}$channels \cite{Valdiviesso2011} and using IceCube data \cite{Esmaili:2014ota} upper limits on $\mathit{\Phi} \Delta\gamma_{21} < 9.1 \times 10^{-27}$ and $\mathit{\Phi} \Delta\gamma_{31} < 6.0 \times 10^{-27}$.

There are other studies too involving neutrinos and probing VEP. It is also claimed that the $\Delta \gamma \simeq 10^{-11}-10^{-13}$ \cite{Weinberg1964} between photons and neutrinos can be probed using time delays between these particles by studying the recently detected 5 TeV neutrinos from Gamma Ray Bursts (GRBs) at IceCube. Similar work has also been studied in \cite{Longo:1987gc,Guzzo:2001vn}. It is being studied and reported that the absence of a superselection rule in Quantum gravity will imply a violation of the Equivalence principle \cite{Adunas:2000zm}; but this effect is much smaller than what we are trying to probe here and doesn't concern our treatment. 

The authors in \cite{Diaz:2020aax} have established the ability of DUNE to probe violations of the equivalence principle (VEP) assuming different gravitational mixing structures. They have also
investigated modified Hamiltonian terms with different energy
dependences and their connection to Lorentz-violating operators in
the Standard Model Extension (SME). In \cite{Calatayud-Cadenillas:2024wdw}, authors have investigated the ability of DUNE to distinguish VEP from other beyond-standard neutrino-oscillation scenarios.In contrast, this work focuses specifically on how the choice of the experimental beam-energy configuration affects the VEP sensitivity at DUNE. We restrict the gravitational
mixing to the $(2,3)$ sector and systematically investigate the ability of  
Low-Energy (LE) \cite{DUNE:2020ypp} and Medium-Energy (ME) \cite{DUNE:2020ypp,Fields:2021duneflux,Rout:2020cxi} beam configurations as DUNE to constrain the  VEP
parameters. We further assess the relative contributions of
neutrino and antineutrino appearance and disappearance channels,
while marginalizing over the standard oscillation parameters,
VEP-sector parameters, and Earth's matter density. Despite the linear
energy dependence of the VEP contribution, we find that the LE
configuration generally provides stronger constraints on the VEP
parameters than the ME configuration.\\

The paper is organized as follows. In Section~2, we outline the theoretical formalism of neutrino propagation in the presence of VEP, incorporating standard matter effects and the parameterization of the exotic gravitational mixing matrix. Section~3 describes the DUNE experimental setup, detailing the baseline parameters and comparing the Low-Energy (LE) and Medium-Energy (ME) flux configurations. Section~4 presents the details of our numerical simulation framework, benchmark scenarios, and standard oscillation parameter settings. In Section~5, we report our findings.

\section{Neutrino propagation in the presence of VEP}
\label{theory}
\hspace*{5ex}
For neutrino propagation over the DUNE baseline, the relevant gravitational field can be
treated as weak and approximately static. In the presence of a possible violation of the
equivalence principle (VEP), different neutrino propagation eigenstates may couple
non-universally to the gravitational field. We parameterize this non-universal coupling as
\begin{equation}
G_N^{(i)}=\gamma_i G_N,
\end{equation}
or, equivalently,
\begin{equation}
\mathit{\Phi}_i=\gamma_i\mathit{\Phi},
\end{equation}
where $\gamma_i$ denotes the gravitational coupling of the $i$-th propagation eigenstate
and $\mathit{\Phi}$ is the ambient gravitational potential.

\hspace*{7ex}
In the weak-field and ultrarelativistic limit, the energy of the $i$-th neutrino
propagation eigenstate can be written as
\begin{equation}
E_i\simeq
E+\frac{m_i^2}{2E}+2E\gamma_i\mathit{\Phi},
\end{equation}
where terms of order $\mathcal{O}(\mathit{\Phi}^2)$ and higher-order terms in
$m_i^2/E^2$ have been neglected. Consequently, the energy difference between
two propagation eigenstates is
\begin{equation}
\Delta E_{ij}
\simeq
\frac{\Delta m_{ij}^2}{2E}
+
2E\mathit{\Phi}\Delta\gamma_{ij},
\end{equation}
where
\begin{equation}
\Delta m_{ij}^2=m_i^2-m_j^2,
\qquad
\Delta\gamma_{ij}=\gamma_i-\gamma_j.
\end{equation}

\hspace*{7ex}
For the DUNE baseline, the neutrino propagation time is of the order of a few
milliseconds. Thus, the variation of the gravitational potential due to the rotation
of the Earth can safely be neglected, and the gravitational potential can be treated
as approximately constant during neutrino propagation. The gravitational potential
at the Earth receives contributions from the Earth, the Sun, the Galaxy, and
large-scale astrophysical structures. For terrestrial neutrino experiments, we
consider the relevant background potential to be approximately constant along the
baseline and denote it simply by $\mathit{\Phi}$.

\hspace*{7ex}
In the flavor basis, the neutrino evolution is governed by the Schr\"odinger equation
\begin{equation}
i\frac{d}{dx}\Psi_F=\mathcal{H}_F\Psi_F,
\end{equation}
where
\begin{equation}
\Psi_F=
\begin{pmatrix}
\nu_e\\
\nu_\mu\\
\nu_\tau
\end{pmatrix}.
\end{equation}

In the standard three-neutrino framework, the effective Hamiltonian in the presence
of standard matter effects and VEP can be written as
\begin{equation}
\mathcal{H}_F=
\frac{1}{2E}
\left[
U_{\rm PMNS}\mathbb{M}^2U_{\rm PMNS}^{\dagger}
+\mathbb{V}
+\mathbb{A}
\right].
\end{equation}

\hspace*{7ex}
After subtracting an irrelevant common contribution from the neutrino mass-squared
matrix, we take
\begin{equation}
\mathbb{M}^2=
\begin{pmatrix}
0&0&0\\
0&\Delta m_{21}^2&0\\
0&0&\Delta m_{31}^2
\end{pmatrix}.
\end{equation}

The standard matter contribution is given by
\begin{equation}
\mathbb{A}=
\begin{pmatrix}
A_{\rm CC}&0&0\\
0&0&0\\
0&0&0
\end{pmatrix},
\end{equation}
where
\begin{equation}
A_{\rm CC}=2EV_{\rm CC}
=2\sqrt{2}EG_FN_e.
\end{equation}

\hspace*{7ex}
The VEP contribution can be expressed in terms of the gravitational mixing matrix
$U_G$, which relates the gravitational propagation eigenstates to the flavor basis.
Since an overall common gravitational contribution does not affect neutrino
oscillations, we subtract $\gamma_1$ from all gravitational couplings and define
\begin{equation}
\Delta\gamma_{21}=\gamma_2-\gamma_1,
\qquad
\Delta\gamma_{31}=\gamma_3-\gamma_1.
\end{equation}

We introduce the dimensionless VEP parameters
\begin{equation}
t_1\equiv\mathit{\Phi}\Delta\gamma_{21},
\qquad
t_2\equiv\mathit{\Phi}\Delta\gamma_{31}.
\end{equation}

The VEP contribution can then be written as
\begin{equation}
\mathbb{V}=
U_G
\begin{pmatrix}
0&0&0\\
0&4E^2t_1&0\\
0&0&4E^2t_2
\end{pmatrix}
U_G^\dagger.
\end{equation}

No absolute value is introduced in these definitions, since the relative signs of
$t_1$ and $t_2$ can be relevant when the VEP contribution interferes with the standard
mass-induced contribution.

\hspace*{7ex}
In general, the gravitational mixing matrix $U_G$ need not be identical to the
standard leptonic mixing matrix $U_{\rm PMNS}$. We consider a gravitational mixing
in the $(2,3)$ sector and parameterize the corresponding mixing matrix as
\begin{equation}
U_G=
\begin{pmatrix}
1&0&0\\
0&\cos\theta_N&\sin\theta_N\\
0&-\sin\theta_N e^{-2i\mathit{\Phi}_N}
&\cos\theta_N e^{-2i\mathit{\Phi}_N}
\end{pmatrix},
\end{equation}
where $\theta_N$ denotes the gravitational mixing angle and $\mathit{\Phi}_N$ is the
corresponding gravitational phase.

\hspace*{7ex}
An equivalent parameterization of the same gravitational mixing matrix, up to
unphysical rephasing, can be written as
\begin{equation}
U_G=
\begin{pmatrix}
1&0&0\\
0&e^{i\mathit{\Phi}_N}&0\\
0&0&e^{-i\mathit{\Phi}_N}
\end{pmatrix}
\begin{pmatrix}
1&0&0\\
0&\cos\theta_N&\sin\theta_N\\
0&-\sin\theta_N&\cos\theta_N
\end{pmatrix}.
\end{equation}

Using this parameterization in the VEP contribution gives
\begin{equation}
\mathbb{V}=
\begin{pmatrix}
0&0&0\\
0&
4E^2\left(
t_1\cos^2\theta_N
+t_2\sin^2\theta_N
\right)
&
4E^2 e^{2i\mathit{\Phi}_N}
(t_2-t_1)
\sin\theta_N\cos\theta_N
\\
0&
4E^2 e^{-2i\mathit{\Phi}_N}
(t_2-t_1)
\sin\theta_N\cos\theta_N
&
4E^2\left(
t_2\cos^2\theta_N
+t_1\sin^2\theta_N
\right)
\end{pmatrix}.
\end{equation}

\hspace*{7ex}
For completeness, the standard leptonic mixing matrix is parameterized as
\begin{equation}
U_{\rm PMNS}=R^{23}W^{13}R^{12},
\end{equation}
where
\begin{equation}
R^{23}=
\begin{pmatrix}
1&0&0\\
0&\cos\theta_{23}&\sin\theta_{23}\\
0&-\sin\theta_{23}&\cos\theta_{23}
\end{pmatrix},
\end{equation}
\begin{equation}
W^{13}=
\begin{pmatrix}
\cos\theta_{13}&0&
\sin\theta_{13}e^{i\delta_{\rm CP}}\\
0&1&0\\
-\sin\theta_{13}e^{-i\delta_{\rm CP}}&0&
\cos\theta_{13}
\end{pmatrix},
\end{equation}
and
\begin{equation}
R^{12}=
\begin{pmatrix}
\cos\theta_{12}&\sin\theta_{12}&0\\
-\sin\theta_{12}&\cos\theta_{12}&0\\
0&0&1
\end{pmatrix}.
\end{equation}

\hspace*{7ex}
Consequently, the complete Hamiltonian governing neutrino propagation in matter in
the presence of VEP is
\begin{equation}
\mathcal{H}_F=
\frac{1}{2E}
\left[
U_{\rm PMNS}
\begin{pmatrix}
0&0&0\\
0&\Delta m_{21}^2&0\\
0&0&\Delta m_{31}^2
\end{pmatrix}
U_{\rm PMNS}^{\dagger}
+
\mathbb{V}
+
\begin{pmatrix}
2\sqrt{2}EG_FN_e&0&0\\
0&0&0\\
0&0&0
\end{pmatrix}
\right].
\end{equation}

\hspace*{7ex}
Thus, the standard mass-induced contribution, the VEP contribution, and the
matter-induced contribution are incorporated consistently in the same flavor-basis
Hamiltonian. In particular, after division by $2E$, the standard mass-induced
contribution scales as $1/E$, whereas the VEP contribution scales as $E$. This
distinct energy dependence provides sensitivity to possible violations of the
equivalence principle in long-baseline neutrino experiments.

\subsection{Analysis of oscillation probabilities}
\label{subsec:oscillation_probability}


\begin{figure}
    \centering
    \includegraphics[width=0.49\linewidth]{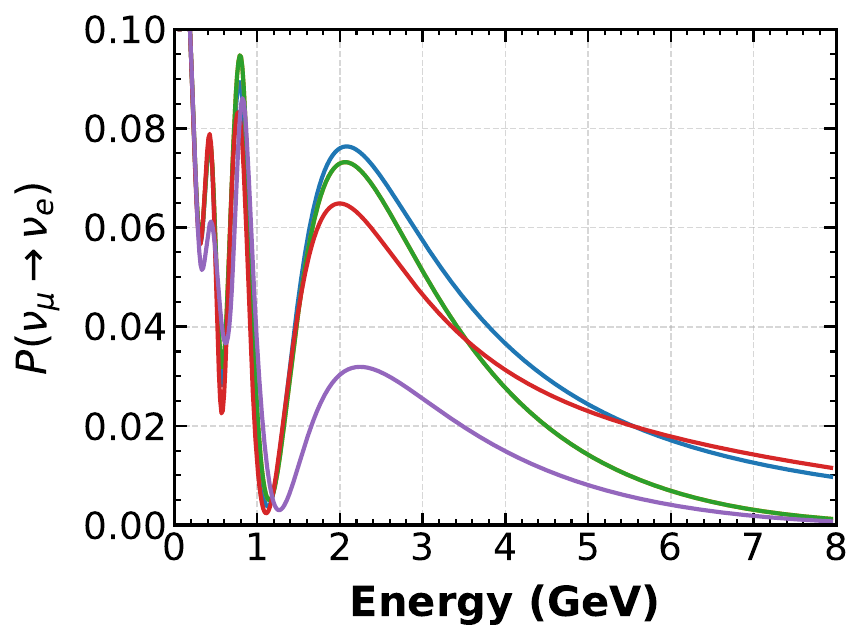}
    \includegraphics[width=0.49\linewidth]{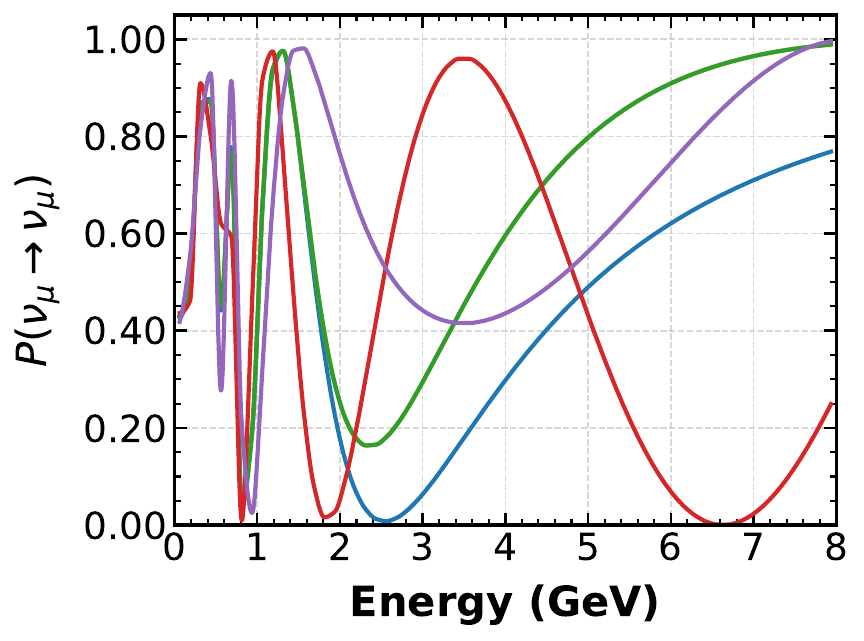}
    \vspace{0.5cm} 
    
\fbox{
\begin{tabular}{ll}
    \textcolor{myblue}{\rule[0.5ex]{1.5em}{1.5pt}} Standard & 
    \textcolor{myorange}{\rule[0.5ex]{1.5em}{1.5pt}} $\theta_N = 0^\circ,\ \phi_N = 0^\circ$ \\
    
    \textcolor{mygreen}{\rule[0.5ex]{1.5em}{1.5pt}} $\theta_N = 0^\circ,\ \phi_N = -90^\circ$ & 
    \textcolor{myred}{\rule[0.5ex]{1.5em}{1.5pt}} $\theta_N = 45^\circ,\ \phi_N = 0^\circ$ \\
    
    \textcolor{mypurple}{\rule[0.5ex]{1.5em}{1.5pt}} $\theta_N = 45^\circ,\ \phi_N = -90^\circ$ & \\
\end{tabular}
}
\caption{Appearance $P(\nu_\mu \rightarrow \nu_e)$ (left) and disappearance $P(\nu_\mu \rightarrow \nu_\mu)$ (right) probabilities as a function of neutrino energy across the 1300\,km DUNE baseline. The standard three-flavor oscillation scenario (blue curve) is compared against four VEP benchmark cases ($t_1 = 5 \times 10^{-23}$, $t_2 = 0$) with varied exotic mixing angles ($\theta_N = 0^\circ, 45^\circ$) and exotic CP phases ($\Phi_N = 0^\circ, -90^\circ$). Note that for $\theta_N = 0^\circ$, the $\Phi_N = 0^\circ$ curve (orange) lies directly underneath the $\Phi_N = -90^\circ$ curve (green) due to phase degeneracy at zero exotic mixing. Standard oscillation parameters are fixed to global best-fit values with $\delta_{\text{CP}} = 212^\circ$.}
    \label{fig:prob_mue}
\end{figure}

Here in Fig.~\ref{fig:prob_mue}, we show the appearance (left panel) and disappearance (right panel) probabilities across the $0\text{--}8\,\mathrm{GeV}$ kinematic window. In the standard three-flavor paradigm (blue curve on the left figure), $P(\nu_\mu \rightarrow \nu_e)$ reaches its first oscillation maximum at $E_\nu \approx 2.1\,\mathrm{GeV}$ with a peak height of $P \approx 0.076$, followed by a sharp kinematic node (minimum) near $E_\nu \approx 1.2\,\mathrm{GeV}$. Below $1.0\,\mathrm{GeV}$, higher-order oscillation cycles appear, peaking at $P \approx 0.095$ near $E_\nu \approx 0.8\,\mathrm{GeV}$.

The introduction of leading-order VEP effects ($t_1 = 5 \times 10^{-23}$) produces distinct spectral modifications depending on the exotic mixing parameters ($\theta_N, \Phi_N$). For a small mixing angle ($\theta_N = 0^\circ$), the exotic phase $\Phi_N$ becomes physically unobservable. As a result, the curve for ($\theta_N = 0^\circ, \Phi_N = 0^\circ$) (orange) is almost entirely overlapped by the curve for ($\theta_N = 0^\circ, \Phi_N = -90^\circ$) (green) across the main oscillation region ($1.0\text{--}4.0\,\mathrm{GeV}$), causing the two benchmark scenarios to appear as a single curve in Fig.~\ref{fig:prob_mue}. Conversely, under maximal mixing ($\theta_N = 45^\circ$), the probability plot exhibits strong sensitivity to the exotic phase $\Phi_N$. For $\Phi_N = 0^\circ$ (red curve), the first oscillation peak drops moderately to $P \approx 0.065$, while for $\Phi_N = -90^\circ$ (purple curve), strong destructive interference suppresses the first maximum down to $P \approx 0.032$.

In the disappearance channel $P(\nu_\mu \rightarrow \nu_\mu)$ (right panel), the standard scenario exhibits its primary oscillation minimum (maximal disappearance) near $E_\nu \approx 2.5\,\mathrm{GeV}$. Consistent with the appearance channel, the small exotic mixing cases ($\theta_N = 0^\circ$) remain phase-degenerate, with the orange and green curves overlapping and shifting the minimum slightly down to $E_\nu \approx 2.2\,\mathrm{GeV}$. Under maximal mixing ($\theta_N = 45^\circ$), the spectral distortions become severe: for $\Phi_N = -90^\circ$ (purple curve), the primary minimum is pushed to a higher energy near $3.2\,\mathrm{GeV}$, while for $\Phi_N = 0^\circ$ (red curve), the oscillation frequency drastically increases, shifting the first minimum down to $E_\nu \approx 1.3\,\mathrm{GeV}$ and creating a second deep node in the high-energy tail near $E_\nu \approx 6.5\,\mathrm{GeV}$.

While these theoretical probability distributions are flux-independent, their experimental observability directly depends on the chosen beam flux mode. The LE beam alignment ($1.0\text{--}4.0\,\mathrm{GeV}$) concentrates event statistics directly around the $2.1\,\mathrm{GeV}$ peak, maximizing sensitivity to the severe amplitude suppression observed in the ($\theta_N = 45^\circ, \Phi_N = -90^\circ$) scenario. On the other hand, the broader ME beam alignment ($3.0\text{--}8.0\,\mathrm{GeV}$) provides statistics across the higher-energy tail, allowing DUNE to probe the spectral splitting between $\Phi_N = 0^\circ$ and $\Phi_N = -90^\circ$ above $4.0\,\mathrm{GeV}$.

\section{Experimental Details}

\textbf{DUNE:} The upcoming Deep Underground Neutrino Experiment (DUNE) in the US \cite{DUNE:2015lol, DUNE:2016hlj, DUNE:2020ypp, DUNE:2020txw} will utilize a high-intensity, broadband neutrino beam generated by the Long-Baseline Neutrino Facility (LBNF) at Fermilab. It will initially operating at 1.2 MW power (with future upgrades planned) \cite{DUNE:2015lol}. The far detector consists of four 10-kton (fiducial) Liquid Argon Time Projection Chambers (LArTPCs) located 1300 km away at the Sanford Underground Research Facility (SURF) \cite{DUNE:2015lol, DUNE:2020txw}. The primary focus of DUNE is to measure neutrino mass ordering, the discovery of leptonic CP violation, and precise measurements of oscillation parameters \cite{DUNE:2015lol, DUNE:2020ypp}; yet, it is designed to be able to investigate physics that extends beyond the Standard Model \cite{DUNE:2020fgq}. The simulation details of DUNE used in this work can be found in \cite{DUNE:2021cuw}. The energy resolution is implemented using a detector-specific smearing matrix which accounts for reconstruction effects.  Throughout this analysis, the matter density is assumed to be constant at $2.848~\text{g/cm}^3$, with a $2\%$ uncertainty. 

\label{subsubsec:neutrino_flux}

\begin{figure}
    \centering
       \includegraphics[width=0.49\linewidth]{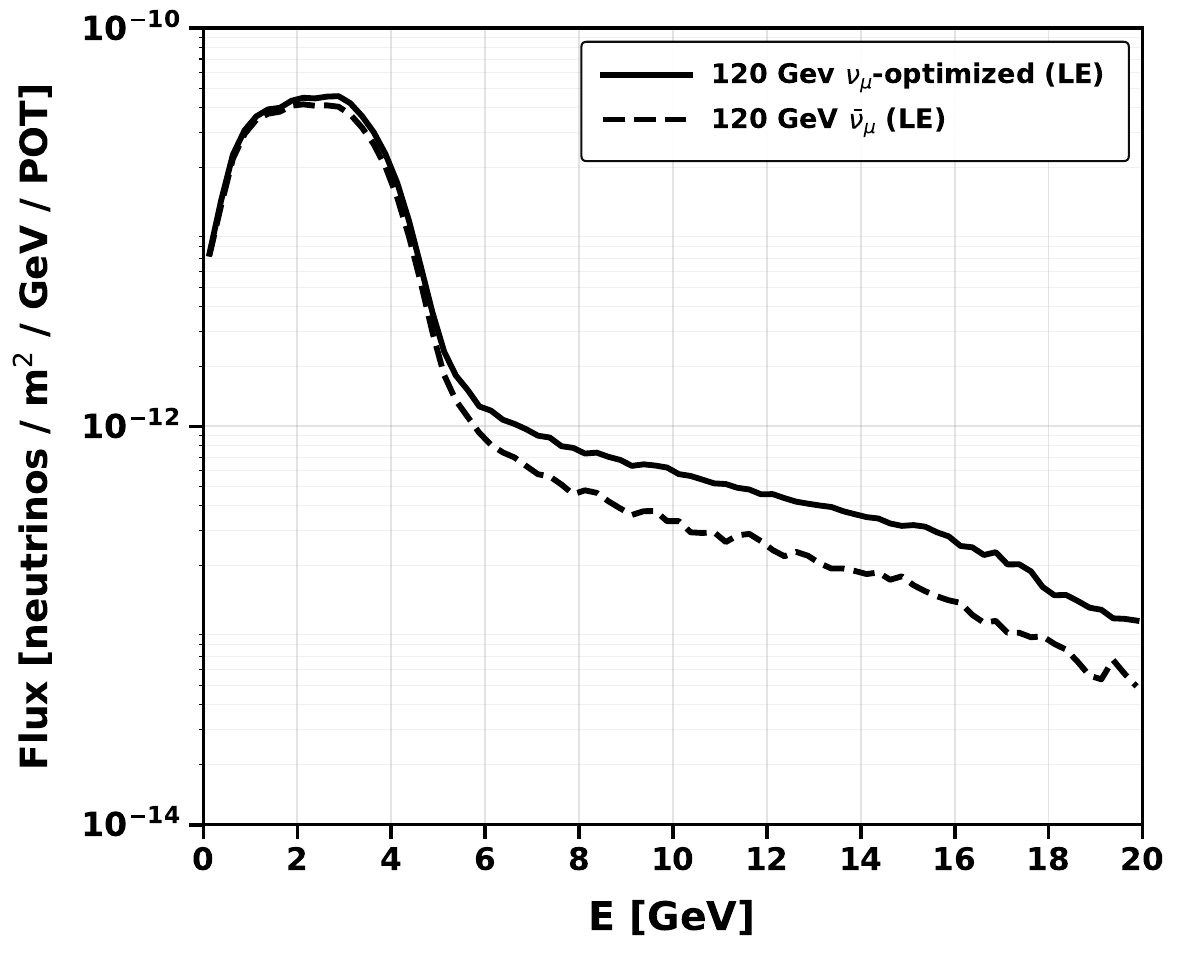}
        \includegraphics[width=0.49\linewidth]{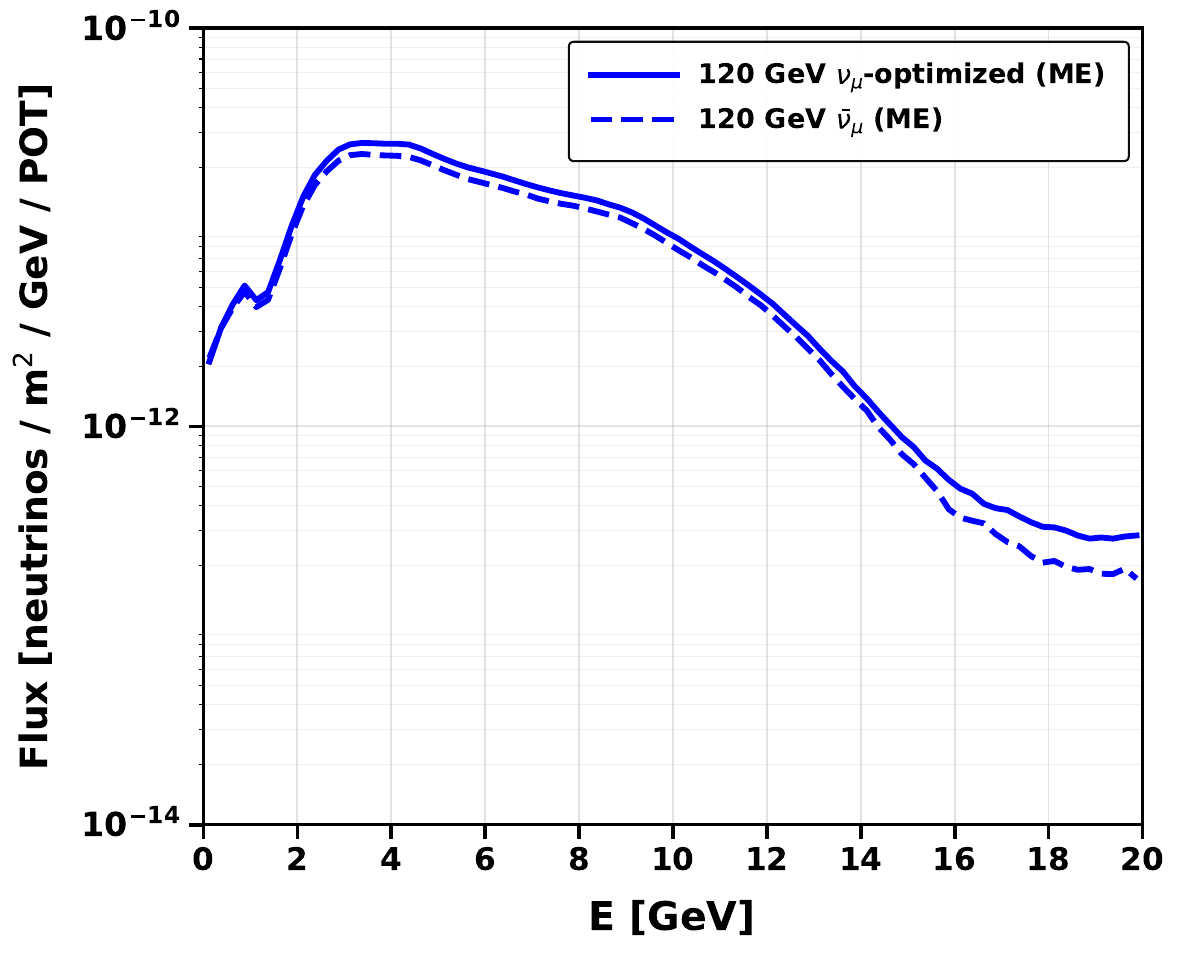}
    \caption{Unoscillated neutrino flux spectra at the DUNE Far Detector baseline ($1300\,\mathrm{km}$) as a function of energy for a $120\,\mathrm{GeV}$ primary proton beam. The left panel shows the Low-Energy (LE) flux configuration \cite{Rout:2020cxi}, and the right panel shows the Medium-Energy (ME) flux configuration \cite{DUNE:2020ypp,Fields:2021duneflux,Rout:2020cxi}. Solid curves represent the $\nu_\mu$ flux in Forward Horn Current (FHC) mode, while dashed curves represent the $\bar{\nu}_\mu$ flux in Reverse Horn Current (RHC) mode. Beam tune setups follow the framework described in Ref.~\cite{Siyeon:2024pte}.}

    \label{fig:flux_le_me}
\end{figure}

The energy-dependent neutrino flux profiles for the Low-Energy (LE) \cite{Rout:2020cxi} and Medium-Energy (ME) beam configurations \cite{DUNE:2020ypp,Fields:2021duneflux,Rout:2020cxi}analyzed in this work are shown in Fig.~\ref{fig:flux_le_me}. Both beam configurations are produced using a $120,\mathrm{GeV}$ primary proton beam impinging on a graphite target, with an assumed exposure of $1.1\times10^{21}$ protons on target (POT) per year, and are simulated using the G4LBNF framework based on GEANT4~\cite{Agostinelli:2002hh, Allison:2006ve}, which models the Long-Baseline Neutrino Facility (LBNF) beamline~\cite{DUNE:2020ypp}. Following the multi-tune beam framework described in Ref.~\cite{Siyeon:2024pte}, these profiles capture the flux distributions delivered to the Far Detector at the $1300,\mathrm{km}$ baseline. The two configurations are considered in two complementary horn-current polarities: Forward Horn Current (FHC), which produces the $\nu_\mu$-optimized neutrino beam (solid curves), and Reverse Horn Current (RHC), which produces the $\bar{\nu}_\mu$-optimized antineutrino beam (dashed curves).

As illustrated in Fig.~\ref{fig:flux_le_me}, the two beam options exhibit distinct spectral features tailored for different kinematic regimes. In the Low-Energy (LE) spectrum, the flux peaks sharply near $E_\nu \approx 2.5\,\mathrm{GeV}$, concentrating maximum intensity directly at the first $\nu_\mu \rightarrow \nu_e$ oscillation maximum for the $1300\,\mathrm{km}$ baseline, while decreasing rapidly above $4.0\,\mathrm{GeV}$ to suppress high-energy neutral-current backgrounds. Conversely, the Medium-Energy (ME) spectrum exhibits a substantially broader flux distribution, peaking between $3.0\,\mathrm{GeV}$ and $5.0\,\mathrm{GeV}$ and maintaining significant intensity up to $10.0\text{--}12.0\,\mathrm{GeV}$, thereby enabling higher event statistics at higher energies and covering sub-leading kinematic regions.

In both LE and ME modes, the overall spectral envelope is preserved across horn polarities, with the RHC ($\bar{\nu}_\mu$) flux magnitude slightly lower than the FHC ($\nu_\mu$) flux due to the lower secondary $\pi^-$ production cross-section relative to $\pi^+$. Employing these complementary LE and ME flux profiles across both horn polarities allows us to evaluate how spectral beam tuning impacts DUNE's sensitivity to VEP-induced modifications.

\section{Details of the simulation}
\label{sec:simulation_details}

For the standard three-flavor oscillation parameters, we adopt central values consistent with global best-fit analyses \cite{Esteban:2024eli}. The mixing angles are fixed to $\theta_{12} = 33.76^\circ$, $\theta_{13} = 8.62^\circ$, and $\theta_{23} = 43.29^\circ$, alongside a Dirac CP-violating phase of $\delta_{\mathrm{CP}} = 212^\circ$. The mass-squared differences are set to $\Delta m_{21}^2 = 7.537 \times 10^{-5}\,\mathrm{eV}^2$ and $\Delta m_{31}^2 = 2.511 \times 10^{-3}\,\mathrm{eV}^2$. Coherent matter effects along the $1300\,\mathrm{km}$ baseline are incorporated assuming a constant Earth matter density of $\rho = 2.848\,\mathrm{g/cm^3}$, serving as the standard line-of-sight approximation for the DUNE trajectory \cite{DUNE:2020ypp}. Throughout this analysis, we assume normal mass ordering (NO) as the true neutrino mass ordering.

To investigate the phenomenological signatures of the Violation of the Equivalence Principle (VEP), exotic parameters are introduced into the effective flavor oscillation Hamiltonian. The VEP mass-coupling parameters are defined as $t_1 = \mathit{\Phi}\Delta\gamma_{21}$ and $t_2 = \mathit{\Phi}\Delta\gamma_{31}$. For the baseline probability and event spectra generation, we set $t_1 = 5 \times 10^{-23}$ and $t_2 = 0$ to isolate the leading-order VEP perturbations.

\section{Event Spectra and Sensitivity Analysis}
\label{sec:results}

In this section, we evaluate the impact of the Violation of the Equivalence Principle (VEP) on neutrino oscillation parameters at DUNE. Phenomenological simulations are executed using the General Long Baseline Experiment Simulator (\texttt{GLoBES}) framework \cite{Huber:2004ka, Huber:2007ji}, incorporating both the Low-Energy (LE) and Medium-Energy (ME) beam tunes. We have modified the neutrino oscillation probabilities implemented in GLoBES according to the theoretical calculations presented in Sec.~\ref{theory}.  To comprehensively evaluate VEP impact across both appearance ($\nu_\mu \rightarrow \nu_e, \bar{\nu}_\mu \rightarrow \bar{\nu}_e$) and disappearance ($\nu_\mu \rightarrow \nu_\mu, \bar{\nu}_\mu \rightarrow \bar{\nu}_\mu$) (See the Appendix) channels, we establish five primary benchmark scenarios: the Standard Oscillation null hypothesis corresponding to pure three-flavor oscillations without VEP effects, Benchmark~I with $\theta_N = 0^\circ$ and $\mathit{\Phi}_N = 0^\circ$, Benchmark~II with $\theta_N = 0^\circ$ and $\mathit{\Phi}_N = -90^\circ$, Benchmark~III with $\theta_N = 45^\circ$ and $\mathit{\Phi}_N = 0^\circ$, and Benchmark~IV with $\theta_N = 45^\circ$ and $\mathit{\Phi}_N = -90^\circ$. These configurations systematically isolate the independent and coupled dependencies on the exotic mixing angle $\theta_N$ and exotic CP phase $\mathit{\Phi}_N$.

\subsection{Event analysis}
\begin{figure}
    \centering
         \includegraphics[width=0.49\linewidth]{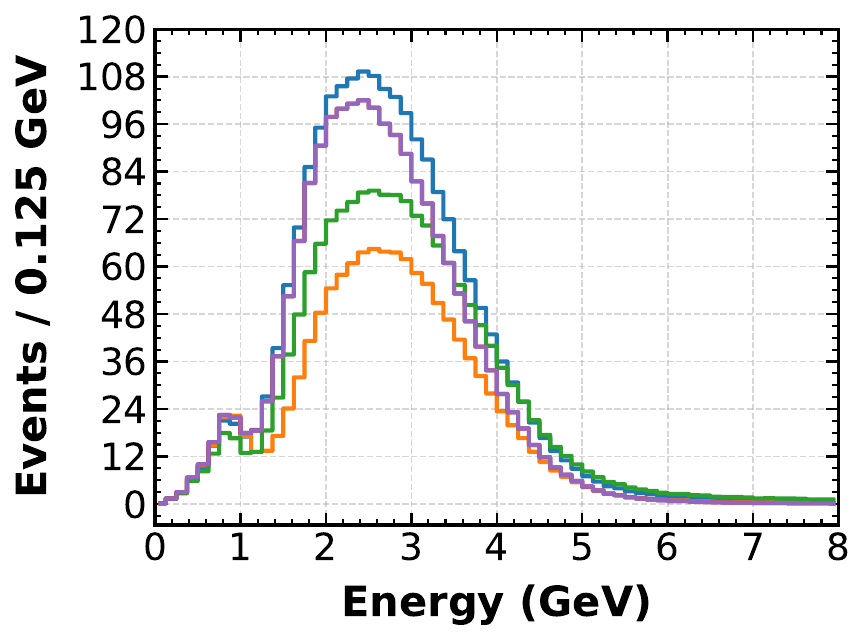}
    \includegraphics[width=0.49\linewidth]{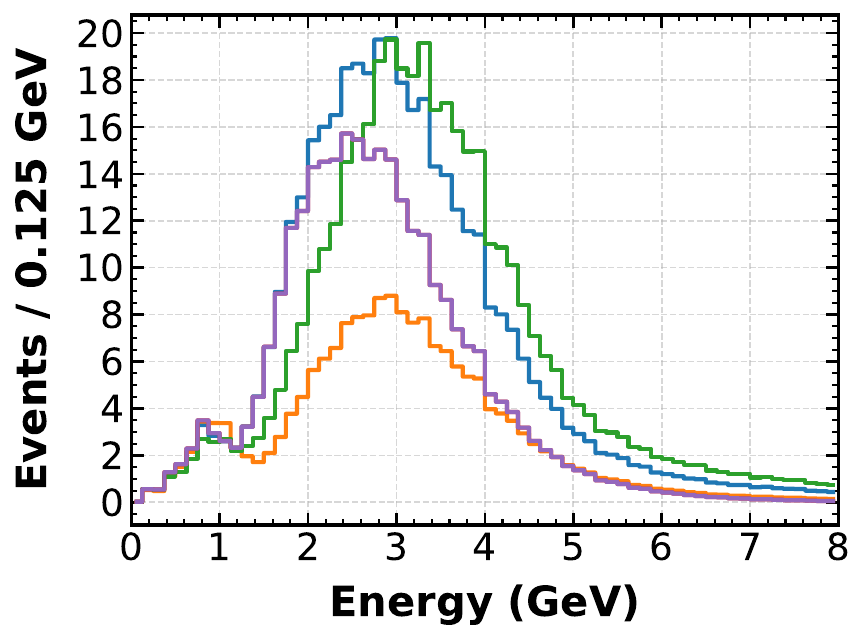}
    \includegraphics[width=0.49\linewidth]{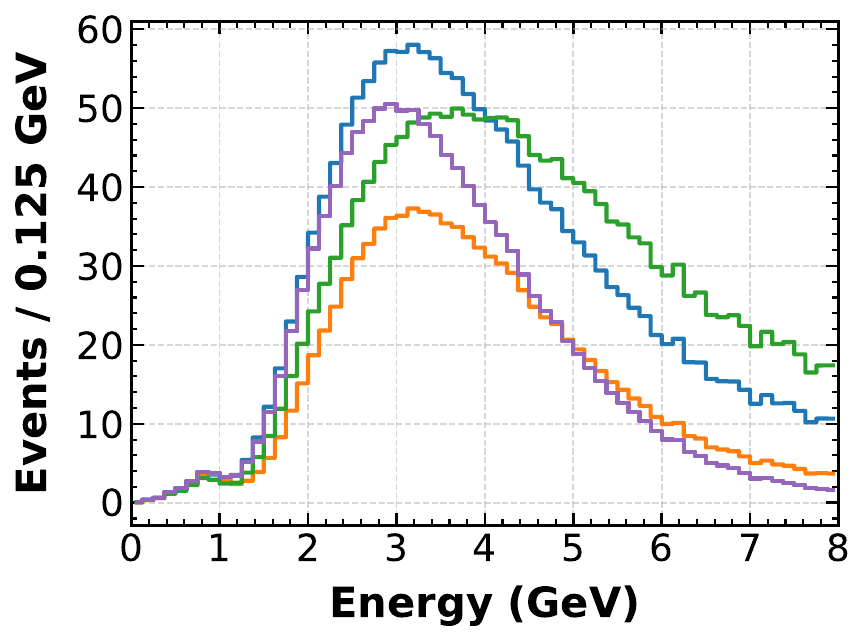}
    \includegraphics[width=0.49\linewidth]{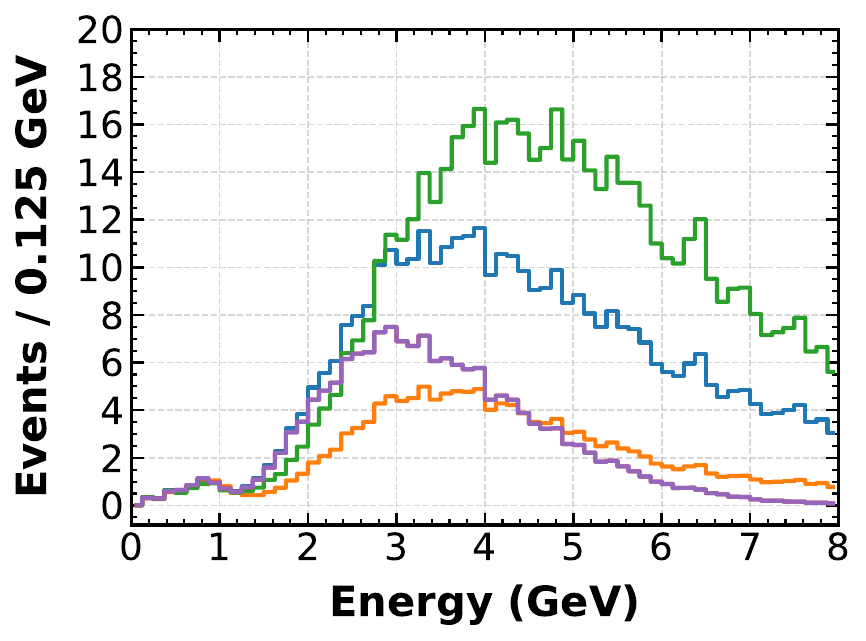}
      \includegraphics[width=0.60\textwidth]{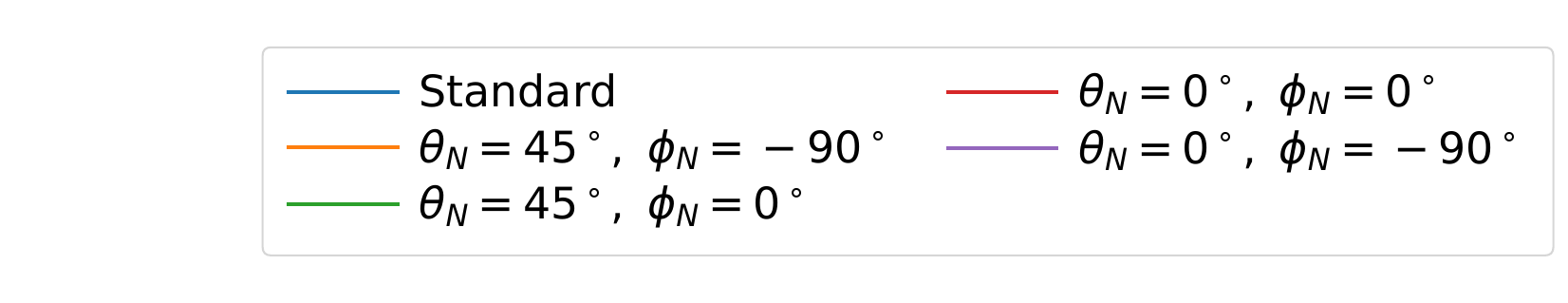}
      
    \caption{Expected reconstructed appearance event spectra at the 40\,kton DUNE Far Detector for a 6.5\,yr FHC + 6.5\,yr RHC run, comparing Low-Energy (top row) and Medium-Energy (bottom row) beam tunes. The top-left (top-right) panel shows the LE $\nu_e$ ($\bar{\nu}_e$) appearance spectrum, while the bottom-left (bottom-right) panel shows the corresponding ME $\nu_e$ ($\bar{\nu}_e$) appearance spectrum, all as functions of the reconstructed neutrino energy. Standard oscillation parameters are fixed to global best-fit values with $\delta_{\text{CP}} = 212^\circ$.}
        \label{fig:appearance_events}.
        
\end{figure}

To establish the primary BSM signatures, we analyse the $\nu_\mu \rightarrow \nu_e$ appearance spectra for the standard three-flavor oscillation scenario (the null hypothesis) alongside, four distinct VEP benchmark cases characterized by varied values of the exotic mixing angle $\theta_N$ and exotic CP-violating phase $\mathit{\Phi}_N$. We fix $t_1 = 5 \times 10^{-23}$ (with $t_2 = 0$) for this analysis. We analyze the reconstructed appearance event spectra ($\nu_e$ and $\bar{\nu}_e$) for both Low-Energy (LE) and Medium-Energy (ME) configurations across a $40\,\mathrm{kton}$ far detector volume over $6.5$ years of operation in each horn polarity ($6.5\,\text{yr FHC} + 6.5\,\text{yr RHC}$, corresponding to a $13\text{--year}$ nominal exposure). Fig~\ref{fig:appearance_events} presents the four appearance datasets in a unified $2\times2$ grid layout. Disappearance channel spectra ($\nu_\mu$ and $\bar{\nu}_\mu$) are deferred to Appendix~\ref{app:disappearance}.




The upper panel of Figs~\ref{fig:appearance_events} display the appearance event distributions for the LE flux profile, the lower panel shows the event spectra for the ME beam configuration. In both the plots in the top panel, we observe that the events are tightly concentrated in the $2.0\text{--}3.0\,\mathrm{GeV}$ region. On the other hand, with the ME configuration (bottom panel), we obtain high statistics up to $8\,\mathrm{GeV}$.

\textbf{\textit{Neutrino mode}}: or the LE configuration, the spectrum for standard three-flavor oscillations (blue curve) peaks around $E_\nu \approx 2.5,\mathrm{GeV}$, whereas that for the ME configuration peaks around $E_\nu \approx 3.2\text{--}3.5,\mathrm{GeV}$. When the exotic mixing angle vanishes ($\theta_N = 0^\circ$), the VEP parameters yield phase-independent spectra for both $\mathit{\Phi}_N = 0^\circ$ (red curve) and $\mathit{\Phi}_N = -90^\circ$ (purple curve). Non-zero exotic mixing ($\theta_N = 45^\circ$) leads to substantial amplitude suppression for $\mathit{\Phi}_N = -90^\circ$ (orange curve), whereas for the CP-conserving case of $\mathit{\Phi}_N = 0^\circ$ (green curve), the peak exhibits a slight shift towards higher energies with higher statistics than $\mathit{\Phi}_N = -90^\circ$. 

\textbf{\textit{Antineutrino mode}}: In antineutrino mode, interestingly, the peaks of the considered VEP scenarios exhibit significant deviations from one another and from the standard three-flavor oscillation case in both the LE and ME configurations. With the ME flux, the number of events noticeably increases compared to the standard case for $\theta_N = 45^\circ$ and $\mathit{\Phi}_N = 0^\circ$.


\subsection{Degeneracy Between Standard and VEP-Induced CP Phases}
\label{subsec:cp_envelopes}

\begin{figure}
    \centering
         \includegraphics[width=0.49\linewidth]{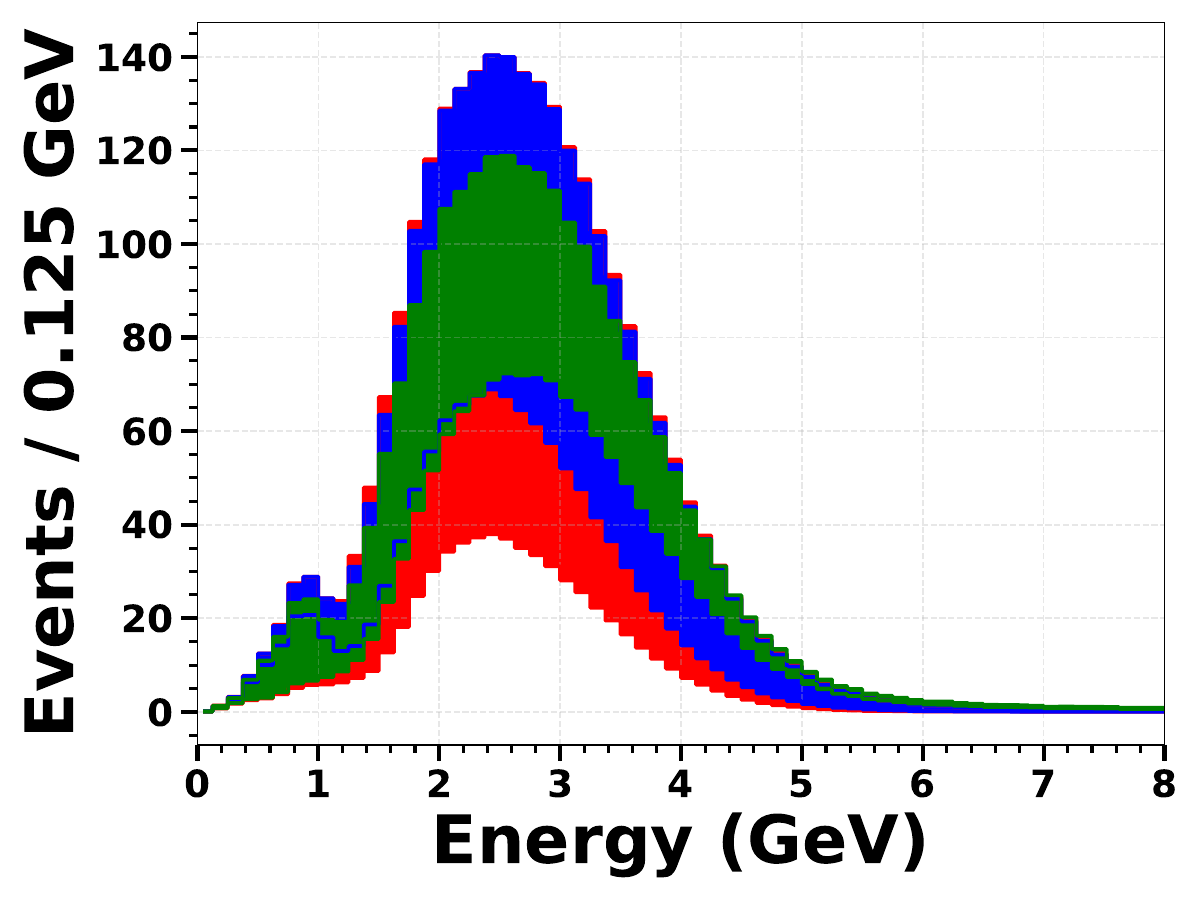}
    \includegraphics[width=0.49\linewidth]{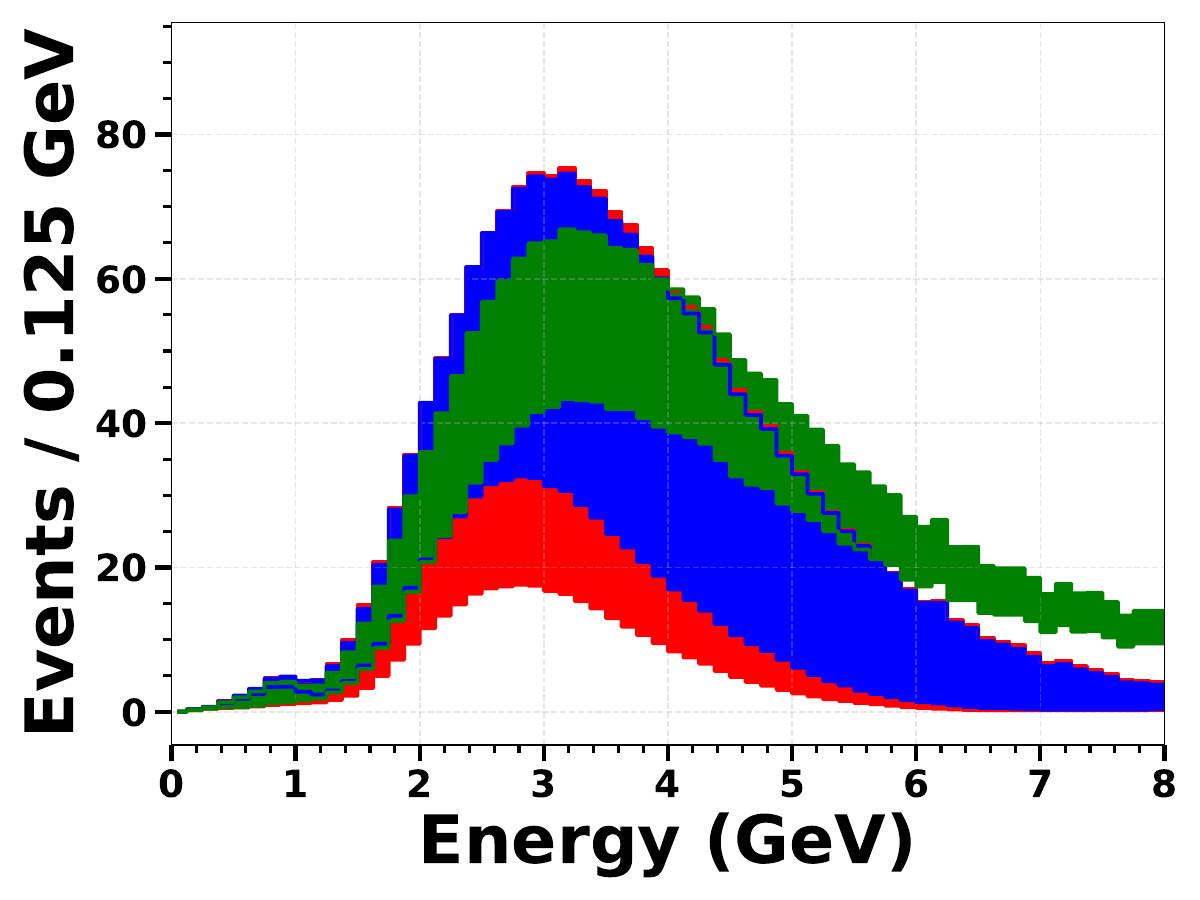}
    
     \vspace{3mm}

    \fbox{%
    \begin{tabular}{@{}ll@{}}
        \textcolor{red}{\rule{1.1ex}{1.1ex}}
        &
        $\boldsymbol{\theta}_{N}=20^\circ,\ 
        \boldsymbol{\delta}_{CP},\mathit{\Phi}_{N}
        \in[-\pi,+\pi]$
        \\[4pt]

        \textcolor{blue}{\rule{1.1ex}{1.1ex}}
        &
        $\boldsymbol{\delta}_{CP}=-90^\circ,\ 
        \boldsymbol{\theta}_{N}=20^\circ,\ 
        \mathit{\Phi}_{N}\in[-\pi,+\pi]$
        \\[4pt]

        \textcolor{green}{\rule{1.1ex}{1.1ex}}
        &
        $\mathbf{Standard:}\ 
        \boldsymbol{\delta}_{CP}\in[-\pi,+\pi]$
    \end{tabular}%
    }
    
\caption{Expected reconstructed appearance event rate envelopes at the $40\,\mathrm{kton}$ DUNE Far Detector for LE (left) and ME (right) flux configurations. The envelopes illustrate event yield ranges when varying: (i) standard CP phase $\delta_{\mathrm{CP}} \in [-\pi, +\pi]$ without VEP (green), (ii) exotic phase $\mathit{\Phi}_N \in [-\pi, +\pi]$ for fixed $\theta_N = 20^\circ$ and $\delta_{\mathrm{CP}} = -90^\circ$ (blue), and (iii) both phases $\delta_{\mathrm{CP}}, \mathit{\Phi}_N \in [-\pi, +\pi]$ simultaneously for $\theta_N = 20^\circ$ (red).}
    \label{fig:cp_event_envelopes}
\end{figure}

Motivated by the distinct spectral features observed for different values of $\mathit{\Phi}_N$ in Fig.~\ref{fig:appearance_events}, we investigate how the standard CP-violating phase $\delta_{\mathrm{CP}}$ and the exotic phase $\mathit{\Phi}_N$ jointly influence the neutrino appearance event spectra. To facilitate this comparison, Fig.~\ref{fig:cp_event_envelopes} shows the minimum and maximum event rates for three specific scenarios. 

The green band (three flavour case) shows the variation in the event spectrum as $\delta_{\mathrm{CP}}$ varies over the range $[-\pi,+\pi]$. The red band represents the variation obtained by simultaneously varying both $\delta_{\mathrm{CP}}$ and $\mathit{\Phi}_N$ over the range $[-\pi,+\pi]$ for $\theta_N = 20^\circ$, while the blue band corresponds to a special case of the red band with $\delta_{\mathrm{CP}} = -90^\circ$.

In the left panel of Fig.~\ref{fig:cp_event_envelopes}, which shows the event spectra for the LE configuration, the green band corresponding to the standard three-flavor oscillation scenario lies entirely within the red band representing the VEP scenario with $\theta_N = 20^\circ$, where both $\delta_{\mathrm{CP}}$ and $\mathit{\Phi}_N$ vary over $[-\pi,+\pi]$. This implies that any event rate predicted by the standard scenario can also be reproduced by an appropriate combination of $\delta_{\mathrm{CP}}$ and $\mathit{\Phi}_N$ in the VEP scenario. Consequently, the LE configuration alone may not be able to distinguish between the standard and VEP scenarios based solely on the event spectra. On the otherhand, in the right panel (ME configuration), we observe that the green and the red band overlap at lower energies but begin to deviate at higher energies. So the degeneracy is broken in the ME configuration at higher energies.

This analysis suggest that the degeneracy between standard CP phase and VEP-induced effects depends on the neutrino energy and the beam configuration. These distinct spectral modifications motivate a quantitative $\chi^2$ analysis to assess the sensitivity of the experiment to VEP effects and determine its ability to distinguish the considered scenarios from standard three-flavor oscillations.

\subsection{\texorpdfstring{$\chi^2$}{chi-square} Analysis}

The statistical sensitivity of the experiment is evaluated by computing the $\Delta\chi^2$ test statistic utilizing the \texttt{GLoBES} framework. To rigorously account for systematic uncertainties, we employ the standard ``pull'' method, incorporating nuisance parameters for both signal and background normalization errors. The Poissonian $\chi^2$ function is constructed as:

\begin{equation}
\chi^2 = \min_{\{\xi_j\}} \left[ \sum_{i,k} 2 \left( N_{i,k}^{th}(\vec{p}, \vec{\xi}) - N_{i,k}^{ex} + N_{i,k}^{ex} \ln \frac{N_{i,k}^{ex}}{N_{i,k}^{th}(\vec{p}, \vec{\xi})} \right) + \sum_{j} \xi_j^2 \right],
\end{equation}

where $N_{i,k}^{ex}$ and $N_{i,k}^{th}$ denote the number of simulated expected events (under the standard oscillation null hypothesis) and the theoretical events (under the VEP hypothesis), respectively, in the $i$-th reconstructed energy bin for the $k$-th interaction channel. The parameter space $\vec{p}$ encompasses the standard oscillation and exotic VEP variables. The set $\{\xi_j\}$ represents the nuisance parameters, with the penalty term $\sum_j \xi_j^2$ constraining them within their prior systematic uncertainties. 

\begin{figure}
    \centering
    \includegraphics[width=0.49\linewidth]{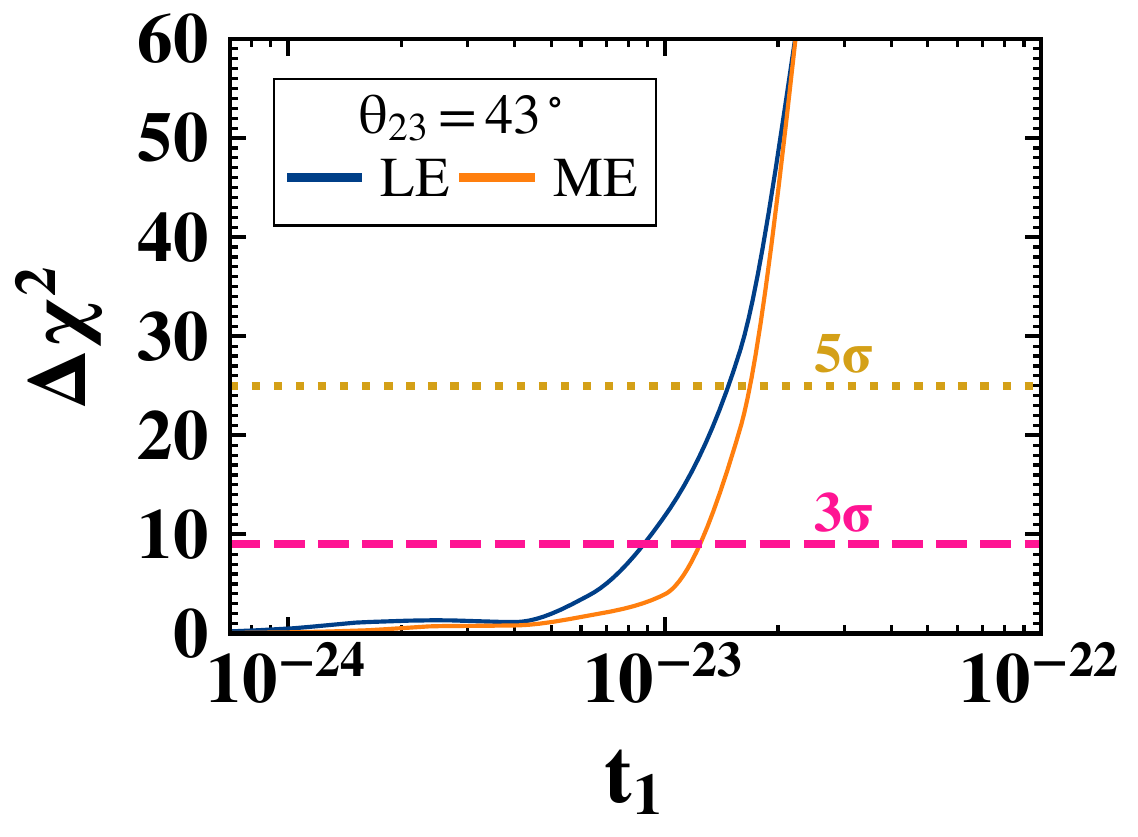}
    \includegraphics[width=0.49\linewidth]{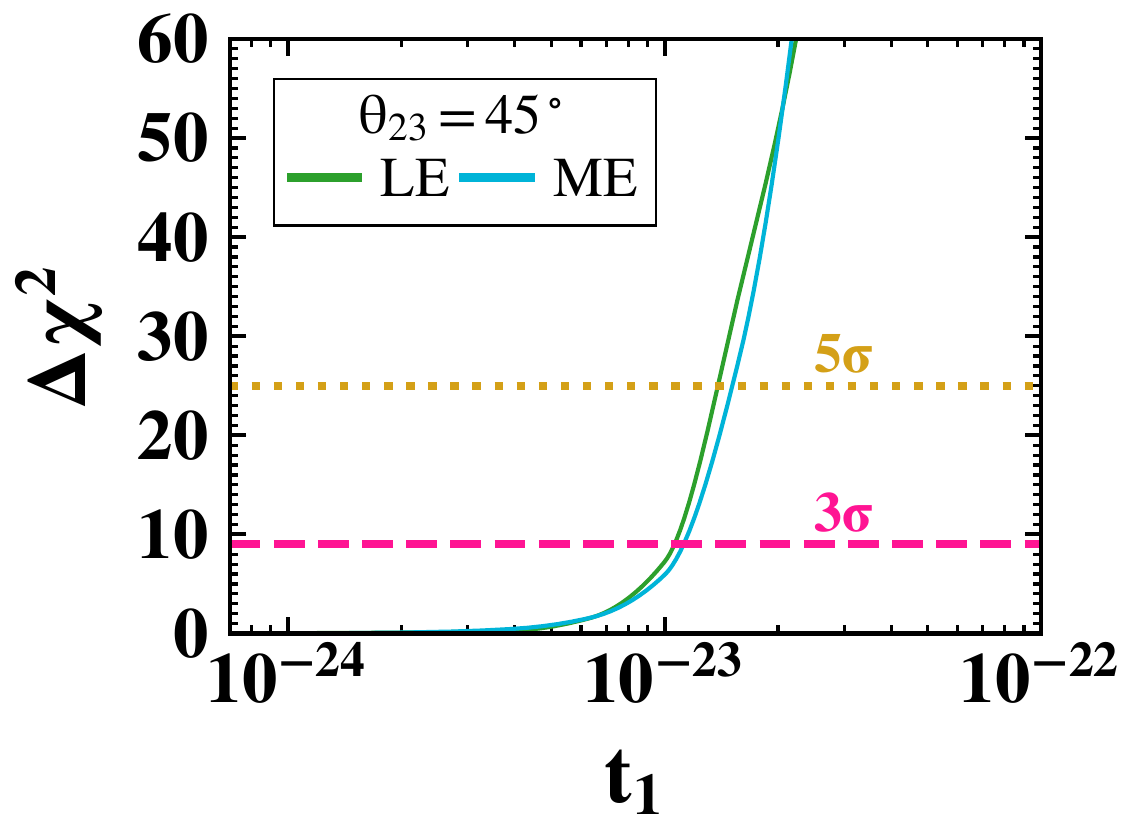}
    \includegraphics[width=0.49\linewidth]{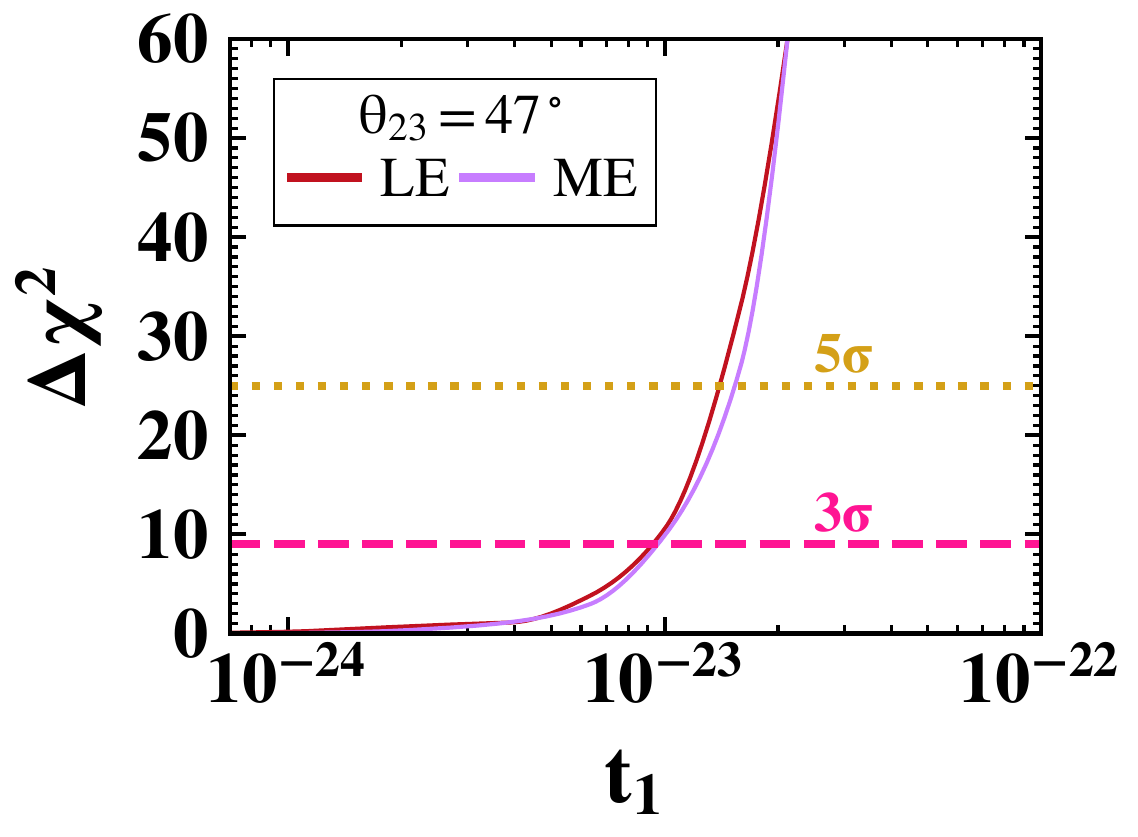}
  
\caption{Expected 1D $\Delta\chi^2$ exclusion plots as a function of the VEP parameter $t_1$, varied over the range $10^{-24}$--$10^{-22}$, for the full DUNE analysis combining appearance and disappearance channels in both FHC and RHC modes. The performance of the Low-Energy (LE) flux configuration is compared with that of the Medium-Energy (ME) configuration for three true values of the atmospheric mixing angle: $\theta_{23}=43^\circ$ (top left), $\theta_{23}=45^\circ$ (top right), and $\theta_{23}=47^\circ$ (bottom). The other VEP parameter is fixed to zero ($t_2 = 0$). }
    \label{fig:chi2_1d_full}
\end{figure}

Throughout this work, the simulated data are generated assuming standard three-flavor neutrino oscillations, while the data are fitted within the framework of VEP-induced new physics. The true values of the standard oscillation parameters are adopted from Ref.~\cite{Esteban:2024eli}. We consider three benchmark values of the atmospheric mixing angle, $\theta_{23}=43^\circ$, $45^\circ$, and $47^\circ$, corresponding to the lower octant, maximal mixing, and upper octant, respectively. For the one-dimensional $\chi^2$ analysis (Figs. ~\ref{fig:chi2_1d_full}, \ref{fig:chi2_1d_full_t2}, \ref{fig:app_disapp_1d} and \ref{fig:nu_anu_1d}), we marginalize over $\delta_{\mathrm{CP}}\in[-\pi,\pi]$ and $\theta_{23}\in[40^\circ,50^\circ]$, while keeping the other standard three-flavor oscillation parameters fixed at their respective best-fit values. The additional mixing angle $\theta_N$ and the VEP-induced phase $\mathit{\Phi}_N$ are also marginalized over the ranges $\theta_N\in[0, 45^\circ]$ and $\mathit{\Phi}_N\in[-\pi,\pi]$, respectively.

To obtain the two-dimensional sensitivity contours, we perform a comprehensive scan over the $\mathit{\Phi}\Delta\gamma_{21}$–$\mathit{\Phi}\Delta\gamma_{31}$ parameter plane, which determines the leading-order VEP parameters $t_1$ and $t_2$. At each point in this plane, the $\chi^2$ function is minimized with respect to all the remaining oscillation and VEP parameters. Specifically, $\delta_{\mathrm{CP}}$ and $\theta_{23}$ are varied over the ranges $[-\pi,\pi]$ and $[40^\circ,50^\circ]$, respectively, while $\theta_{12}$, $\theta_{13}$, $\Delta m^2_{21}$, and $\Delta m^2_{31}$ are varied within their respective $1\sigma$ allowed ranges as reported in Ref.~\cite{Esteban:2024eli}. Additionally, the VEP mixing angle $\theta_N$ and the associated phase $\mathit{\Phi}_N$ are varied over the ranges $[0,\pi/2]$ and $[-\pi,\pi]$, respectively. The minimum $\chi^2$ values obtained at each point are then used to construct the two-dimensional sensitivity contours in the $\mathit{\Phi}\Delta\gamma_{21}$–$\mathit{\Phi}\Delta\gamma_{31}$ parameter plane.


\subsection{Constraining VEP parameters: LE vs. ME comparison}

We first present the comprehensive sensitivity of DUNE by simultaneously combining all four operational interaction channels: $\nu_e$ appearance, $\bar{\nu}_e$ appearance, $\nu_\mu$ disappearance, and $\bar{\nu}_\mu$ disappearance. This full-capacity analysis provides the most rigorous constraints on the fundamental VEP parameters by exploiting statistical power and parameter synergies across all available oscillation modes. 

Computationally, this combined framework is implemented by evaluating the complete set of simulation rules within the customized \texttt{GLoBES} file structure and minimizing the combined $\Delta\chi^2$ contributions from both beam polarities, FHC and RHC, while marginalizing over all other parameters. By evaluating the combined dataset, we systematically analyze the performance of the full experimental configuration under different flux profiles, enabling a direct comparison between the Low-Energy (LE) and Medium-Energy (ME) beam configurations.

\subsubsection{One-Dimensional Sensitivity Analysis}

Fig~\ref{fig:chi2_1d_full} illustrates the $\Delta\chi^2$ test statistic as a function of the VEP parameter $t_1$, varied over the range $10^{-24}$--$10^{-22}$. To ensure a conservative assessment in these one-dimensional projections, the other VEP parameter was fixed to zero ($t_2 = 0$), while the marginalization is done as explain in the previous section. The horizontal reference lines denote the statistical milestones for a $3\sigma$ exclusion ($\Delta\chi^2 = 9$) and a $5\sigma$ expected exclusion sensitivity ($\Delta\chi^2 = 25$).

The one-dimensional profiles demonstrate a strong dependence on the atmospheric mixing octant when comparing the LE and ME beam options. In the lower octant ($\theta_{23} = 43^\circ$, top-left panel), the LE flux exhibits a notably steeper $\Delta\chi^2$ gradient, achieving a $3\sigma$ bound at $8.47 \times 10^{-24}$ compared to $1.15 \times 10^{-23}$ for the ME beam. For maximal mixing ($\theta_{23} = 45^\circ$, top-right panel) and upper octant scenarios ($\theta_{23} = 47^\circ$, bottom panel), the sensitivity gap narrows, with both beam choices providing comparable exclusion bounds near $10^{-23}$ at $3\sigma$ and $1.3\text{--}1.5 \times 10^{-23}$ at $5\sigma$. The exact numerical limits for each octant and flux configuration are summarized in Table~\ref{tab:1d_full_limits}. We observe that the exclusion limit depend upon the choices of the atmospheric mixing angle and for ($\theta_{23} = 43^\circ$, LE configuration can constrain $t_1$ better than ME configuration at 3$\sigma$.

\begin{table}[htbp]
\centering
\caption{Expected upper limits on the VEP parameter
$\mathit{\Phi}\Delta\gamma_{21}$ at the full-capacity DUNE exposure,
obtained from the $\Delta\chi^2$ analysis for different assumed
values of $\theta_{23}$. The limits are quoted at $3\sigma$
($\Delta\chi^2=9$) and $5\sigma$ ($\Delta\chi^2=25$).}
\label{tab:1d_full_limits}
\renewcommand{\arraystretch}{1.2}
\begin{tabular}{c c c c c}
\hline\hline
$\theta_{23}$ &
\multicolumn{2}{c}{$3\sigma$ Upper Limit} &
\multicolumn{2}{c}{$5\sigma$ Upper Limit} \\
\cline{2-3}
\cline{4-5}
&
LE & ME & LE & ME \\
\hline
$43^\circ$ &
$8.47\times10^{-24}$ &
$1.15\times10^{-23}$ &
$1.44\times10^{-23}$ &
$1.64\times10^{-23}$ \\

$45^\circ$ &
$1.03\times10^{-23}$ &
$1.07\times10^{-23}$ &
$1.35\times10^{-23}$ &
$1.49\times10^{-23}$ \\

$47^\circ$ &
$8.98\times10^{-24}$ &
$9.40\times10^{-24}$ &
$1.35\times10^{-23}$ &
$1.51\times10^{-23}$ \\
\hline\hline
\end{tabular}
\end{table}

\begin{figure}
    \centering
    \includegraphics[width=0.49\linewidth]{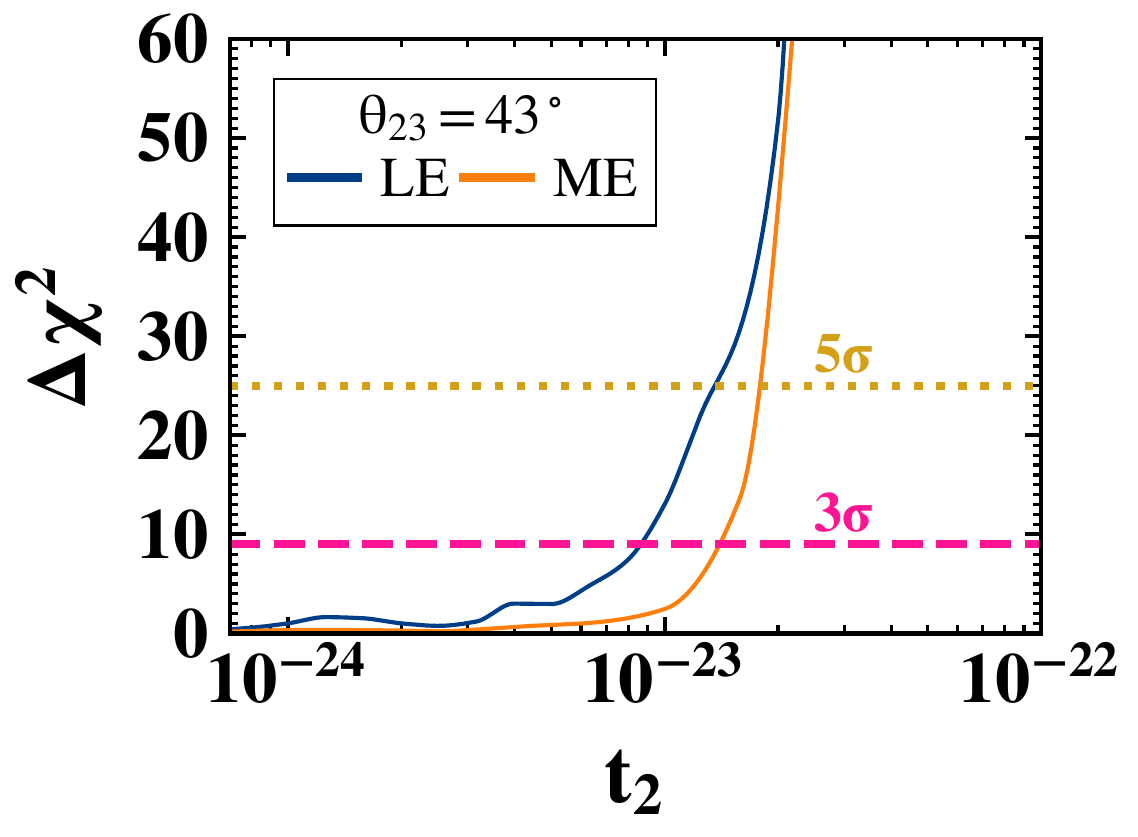}
    \includegraphics[width=0.49\linewidth]{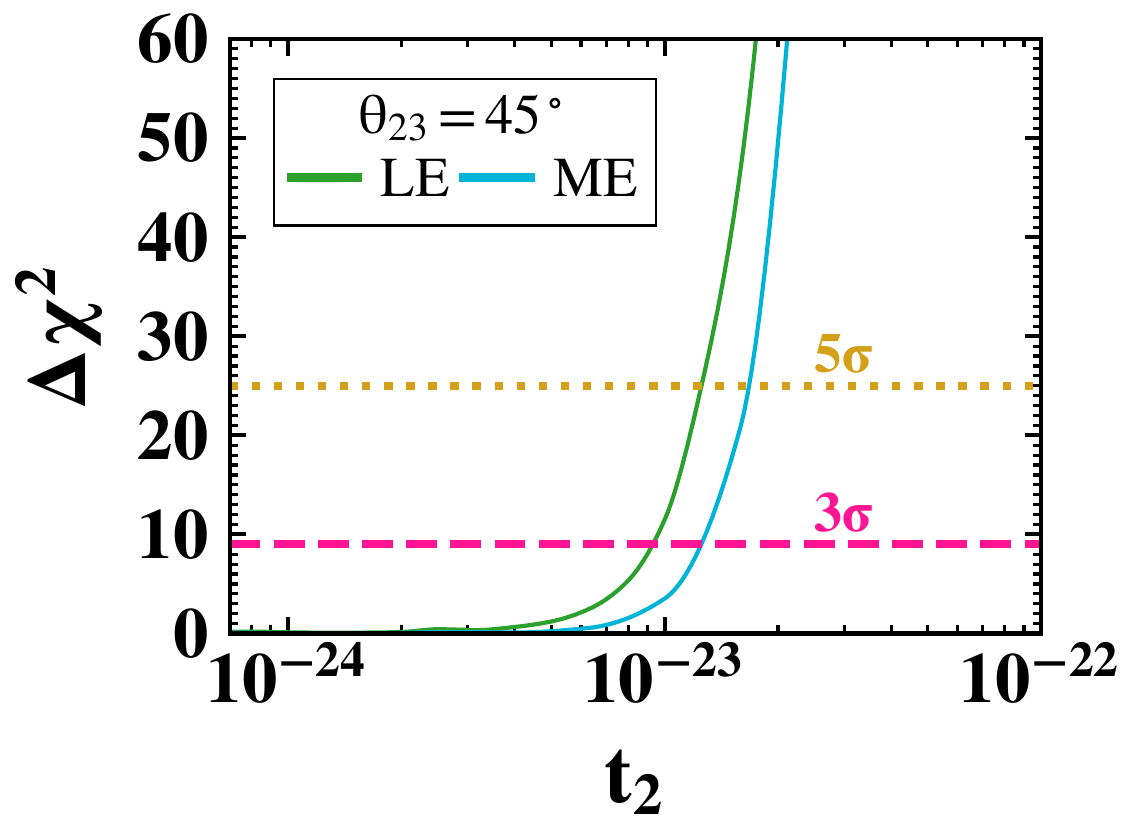}
    \includegraphics[width=0.49\linewidth]{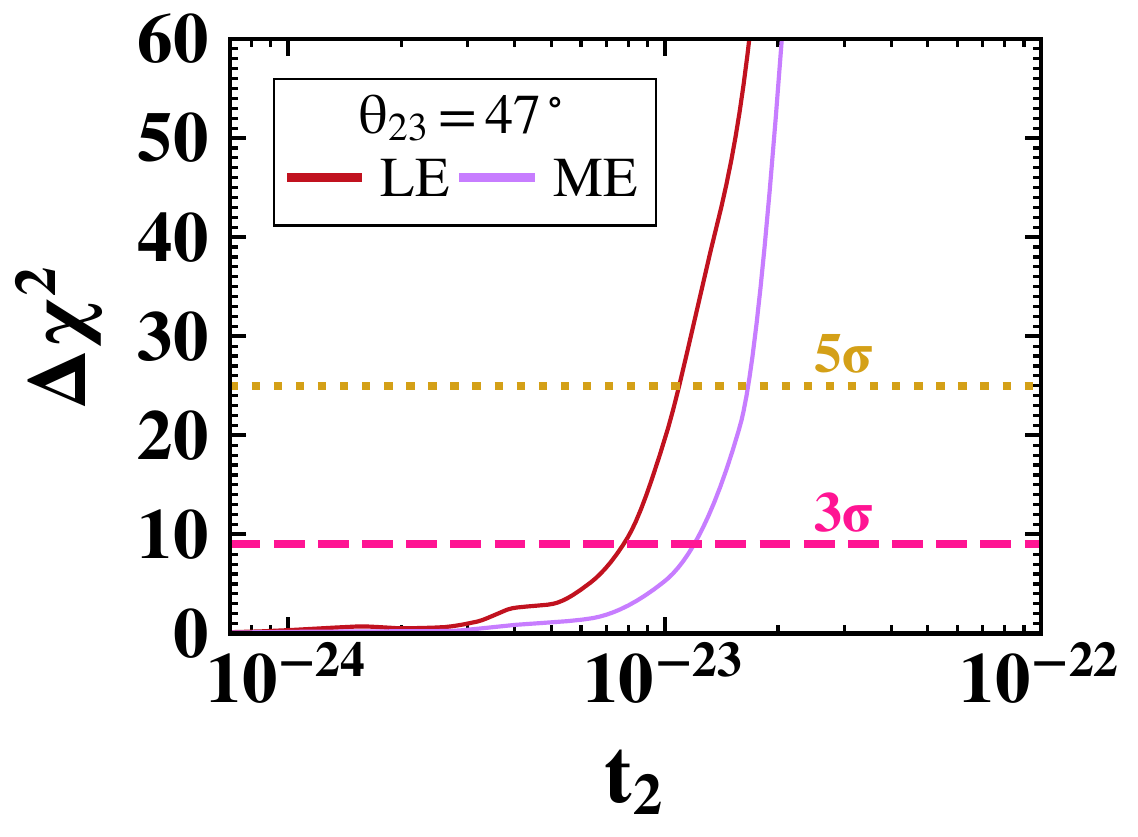}
\caption{Expected 1D $\Delta\chi^2$ exclusion plots as a function of the VEP parameter $t_2$, varied over the range $10^{-24}$--$10^{-22}$, for the full DUNE analysis combining appearance and disappearance channels in both FHC and RHC modes. The performance of the Low-Energy (LE) flux configuration is compared with that of the Medium-Energy (ME) configuration for three true values of the atmospheric mixing angle: $\theta_{23}=43^\circ$ (top left), $\theta_{23}=45^\circ$ (top right), and $\theta_{23}=47^\circ$ (bottom). The other VEP parameter is fixed to zero ($t_1 = 0$).}
    \label{fig:chi2_1d_full_t2}
\end{figure}

Fig.~\ref{fig:chi2_1d_full_t2} presents the corresponding one-dimensional $\Delta\chi^2$ exclusion plot for the other VEP parameter $t_2$, evaluated with $t_1 = 0$. In contrast to the $t_1$ analysis, where the LE and ME configurations performed similarly in the upper octant and maximal mixing scenarios, the $t_2$ constraints demonstrate a strict and uniform advantage for the LE flux configuration across all tested values of $\theta_{23}$. In the lower octant ($\theta_{23} = 43^\circ$), the LE beam achieves a $3\sigma$ exclusion bound of $8.53 \times 10^{-24}$, compared to $1.34 \times 10^{-23}$ for the ME beam. This superiority persists through maximal mixing ($\theta_{23} = 45^\circ$), where LE achieves a $3\sigma$ limit of $9.18 \times 10^{-24}$ compared to ME's $1.19 \times 10^{-23}$. For the upper octant ($\theta_{23} = 47^\circ$), LE achieves $7.73 \times 10^{-24}$ compared to ME's $1.14 \times 10^{-23}$. The LE configuration continues to provide tighter bounds at the $5\sigma$ confidence level as well. The numerical limits for the $t_2$ parameter are detailed in Table~\ref{tab:1d_full_limits_t2}. This persistent LE dominance suggests that the spectral distortions induced primarily by $t_2$ fall predominantly within the $1.0\text{--}4.0\,\mathrm{GeV}$ energy window optimized for the LE flux.

\begin{table}[htbp]
\centering
\caption{Expected upper limits on the VEP parameter
$\mathit{\Phi}\Delta\gamma_{31}$ at the full-capacity DUNE exposure,
obtained from the $\Delta\chi^2$ analysis for different assumed
values of $\theta_{23}$. The limits are quoted at $3\sigma$
($\Delta\chi^2=9$) and $5\sigma$ ($\Delta\chi^2=25$).}
\label{tab:1d_full_limits_t2}
\renewcommand{\arraystretch}{1.2}
\begin{tabular}{c c c c c}
\hline\hline
$\theta_{23}$ &
\multicolumn{2}{c}{$3\sigma$ Upper Limit} &
\multicolumn{2}{c}{$5\sigma$ Upper Limit} \\
\cline{2-3}
\cline{4-5}
&
LE & ME & LE & ME \\
\hline
$43^\circ$ &
$8.53\times10^{-24}$ &
$1.34\times10^{-23}$ &
$1.36\times10^{-23}$ &
$1.73\times10^{-23}$ \\

$45^\circ$ &
$9.18\times10^{-24}$ &
$1.19\times10^{-23}$ &
$1.25\times10^{-23}$ &
$1.64\times10^{-23}$ \\

$47^\circ$ &
$7.73\times10^{-24}$ &
$1.14\times10^{-23}$ &
$1.09\times10^{-23}$ &
$1.63\times10^{-23}$ \\
\hline\hline
\end{tabular}
\end{table}

\subsubsection{2D Contour Analysis}

\begin{figure}
    \centering
    \includegraphics[width=0.49\linewidth]{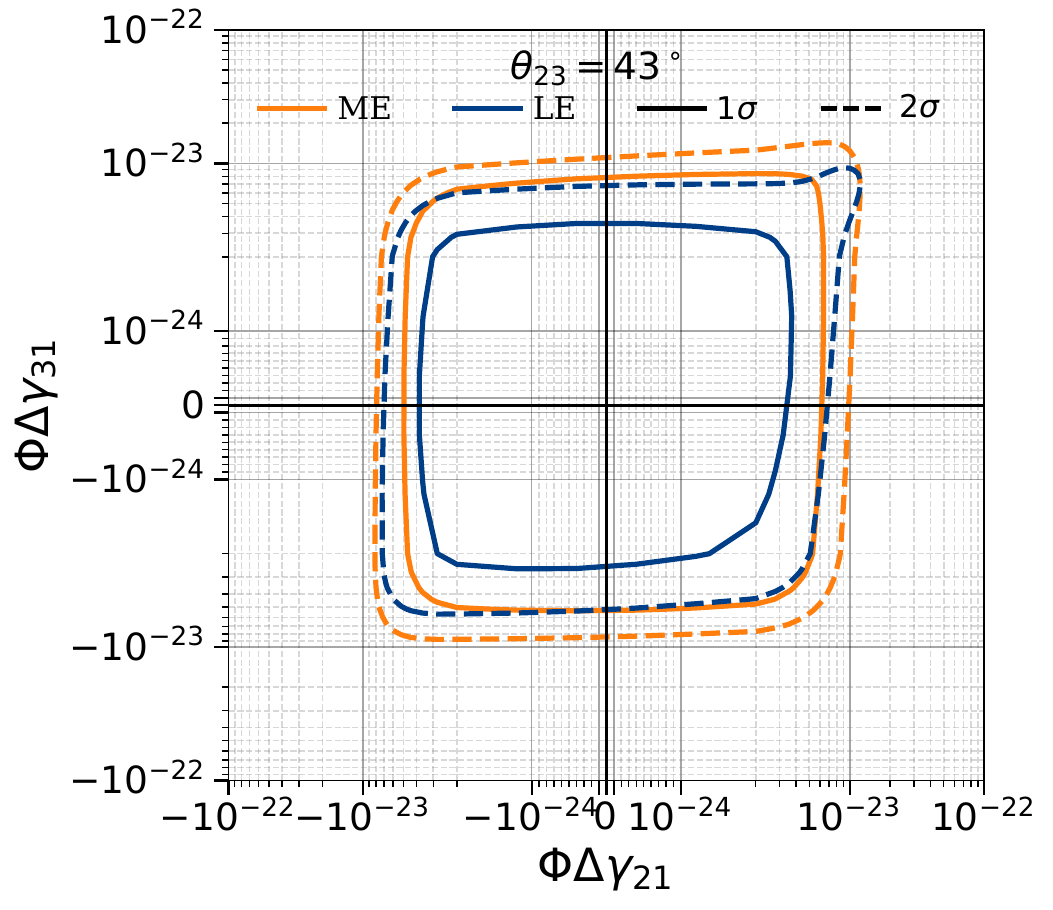}
     \includegraphics[width=0.49\linewidth]{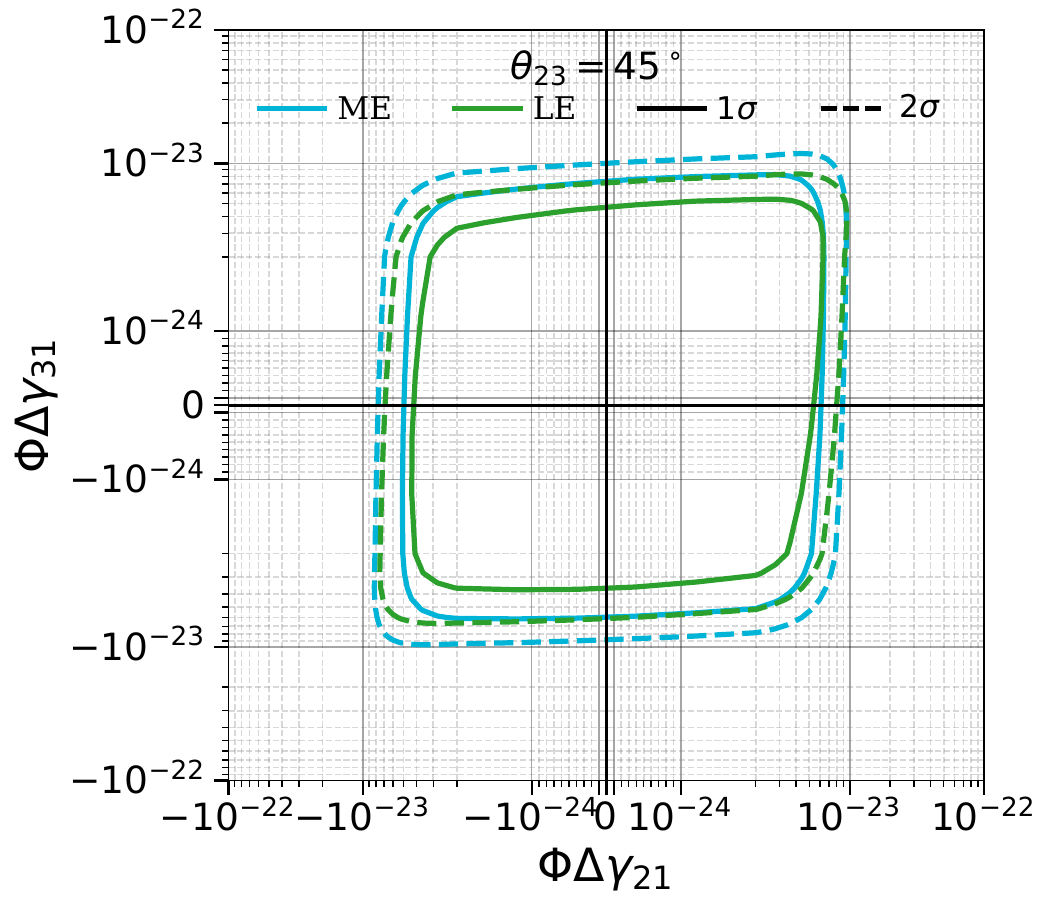}
      \includegraphics[width=0.49\linewidth]{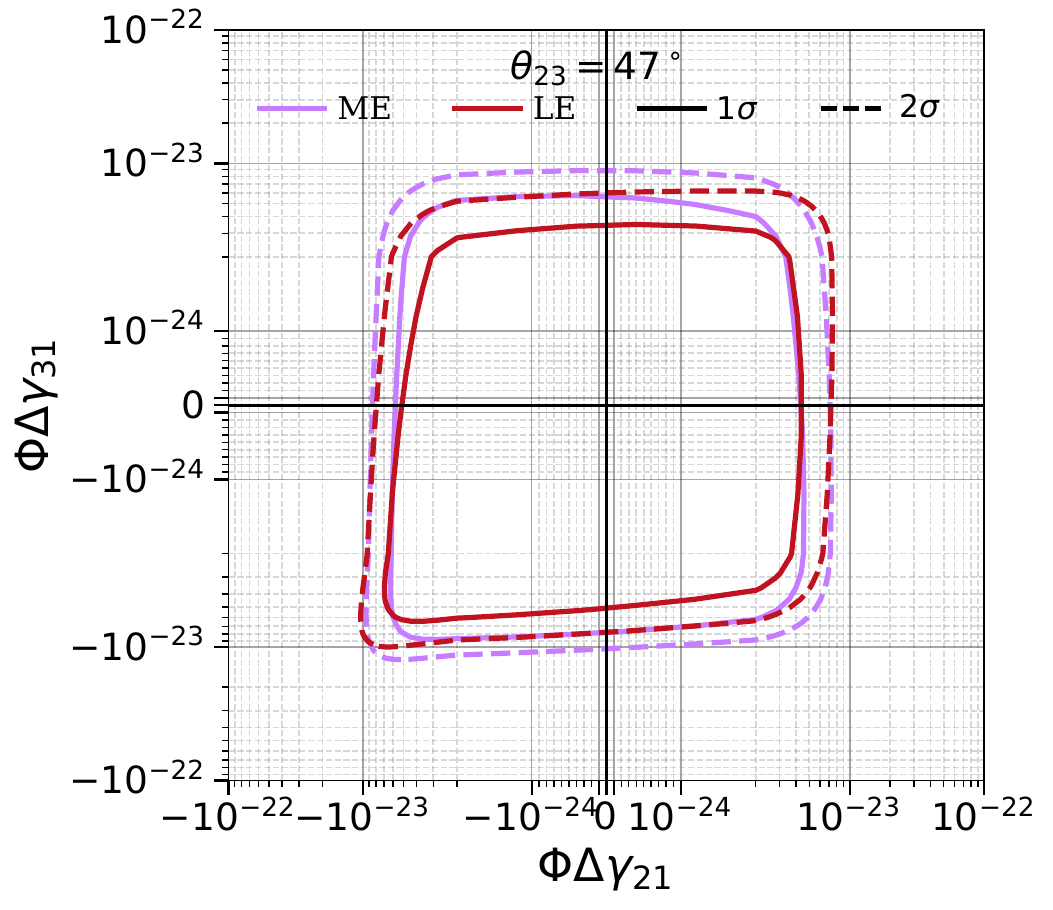}
     
\caption{Expected exclusion contours in the $\mathit{\Phi}\Delta\gamma_{21}$--$\mathit{\Phi}\Delta\gamma_{31}$ plane for the full DUNE analysis, shown for three values of the atmospheric mixing angle: $\theta_{23}=43^\circ$ (top-left), $45^\circ$ (top-right), and $47^\circ$ (bottom). The contours correspond to $1\sigma$ (solid) and $2\sigma$ (dashed) confidence regions for 2 d.o.f. The Low-Energy (LE) and Medium-Energy (ME) beam configurations are represented by different color pairs in each panel. Limits are obtained after marginalizing over standard oscillation parameters, as well as the unconstrained VEP mixing angle $\theta_N$ and phase $\mathit{\Phi}_N$ as mentioned in section \ref{subsec:cp_envelopes}.}
    \label{fig:contour_full}
\end{figure}

To further quantify the experimental sensitivity and evaluate potential correlations between the leading-order VEP mass-coupling parameters, we project the constraints onto the two-dimensional $\Phi\Delta\gamma_{21} - \Phi\Delta\gamma_{31}$ parameter space. Fig.~\ref{fig:contour_full} displays the expected $1\sigma$ and $2\sigma$ confidence level allowed regions for both the LE and ME beam configurations, evaluated across three distinct true values of the atmospheric mixing angle.

 As observed in Fig.~\ref{fig:contour_full}, the two-dimensional allowed regions conclusively reinforce the hierarchy established in the one-dimensional analysis. The contours generated by the LE flux configuration (inner curves) are strictly enveloped by those of the ME configuration (outer curves) across all evaluated scenarios. This geometric nesting physically demonstrates that the narrower, optimized LE beam provides a substantially tighter simultaneous constraint on the full VEP parameter space than the broader ME spectrum.

The structural topology of the sensitivity regions is roughly rectangular with rounded corners, indicating a relatively weak correlation between the leading-order parameters $\mathit{\Phi}\Delta\gamma_{21}$ and $\mathit{\Phi}\Delta\gamma_{31}$ at these confidence levels. While the overall superiority of the LE configuration is invariant, the precise boundary limits exhibit a visible dependence on the true atmospheric octant. At $1\sigma$, for $\theta_{23}=43^\circ$, the approximated allowed ranges for the LE configuration are
$-6.23\times10^{-24}\lesssim\Phi\Delta\gamma_{21}\lesssim5.51\times10^{-24}$
and
$-4.31\times10^{-24}\lesssim\Phi\Delta\gamma_{31}\lesssim4.95\times10^{-24}$,
while for the ME configuration, the corresponding ranges are
$-6.29\times10^{-24}\lesssim\Phi\Delta\gamma_{21}\lesssim7.28\times10^{-24}$
and
$-6.58\times10^{-24}\lesssim\Phi\Delta\gamma_{31}\lesssim1.02\times10^{-23}$.
 The LE configuration provides tighter constraints on both VEP parameters than the ME configuration, irrespective of the true value of $\theta_{23}$. Furthermore, among the three values considered, $\theta_{23}=43^\circ$ yields the smallest allowed region enclosed by the joint confidence contour. Since a smaller allowed region indicates stronger constraints on the correlated parameter space, we conclude that $\theta_{23}=43^\circ$ provides the tightest joint constraint on $\mathit{\Phi}\Delta\gamma_{21}$ and $\mathit{\Phi}\Delta\gamma_{31}$ among the three benchmark values considered. We tabulate the 1$\sigma$ bounds in Table \ref{tab:vep_1sigma_bounds}.
\begin{table*}[htbp]
\centering
\caption{Approximate projected ranges of the VEP parameters
from the $1\sigma$ contours for different values of
$\theta_{23}$.}
\label{tab:vep_1sigma_bounds}
\begin{tabular}{ccccc}
\hline
$\theta_{23}$ & Configuration
& $\Phi\Delta\gamma_{21}$
& $\Phi\Delta\gamma_{31}$ \\
\hline
$43^\circ$ & LE
& $[-6.23 \times 10^{-24},\, 5.51 \times 10^{-24}]$
& $[-4.31 \times 10^{-24},\, 4.95 \times 10^{-24}]$ \\
$43^\circ$ & ME
& $[-6.29 \times 10^{-24},\, 7.28 \times 10^{-24}]$
& $[-6.58 \times 10^{-24},\, 1.02 \times 10^{-23}]$ \\
\hline
$45^\circ$ & LE
& $[-6.08 \times 10^{-24},\, 7.78 \times 10^{-24}]$
& $[-4.86 \times 10^{-24},\, 6.72 \times 10^{-24}]$ \\
$45^\circ$ & ME
& $[-6.22 \times 10^{-24},\, 7.28 \times 10^{-24}]$
& $[-7.25 \times 10^{-24},\, 1.00 \times 10^{-23}]$ \\
\hline
$47^\circ$ & LE
& $[-8.69 \times 10^{-24},\, 6.28 \times 10^{-24}]$
& $[-7.69 \times 10^{-24},\, 4.81 \times 10^{-24}]$ \\
$47^\circ$ & ME
& $[-6.94 \times 10^{-24},\, 5.93 \times 10^{-24}]$
& $[-1.04 \times 10^{-23},\, 7.36 \times 10^{-24}]$ \\
\hline
\end{tabular}
\end{table*}

\subsection{Individual Channel Exclusion Sensitivities}

To understand the origin of the stringent constraints in the full DUNE analysis, we isolate the contributions of the appearance ($\nu_\mu \rightarrow \nu_e, \bar{\nu}_\mu \rightarrow \bar{\nu}_e$) and disappearance ($\nu_\mu \rightarrow \nu_\mu, \bar{\nu}_\mu \rightarrow \bar{\nu}_\mu$) channels. We perform this analysis only for $t_1$ assuming $t_2 = 0$.

\begin{figure}
    \centering
    \includegraphics[width=0.49\linewidth]{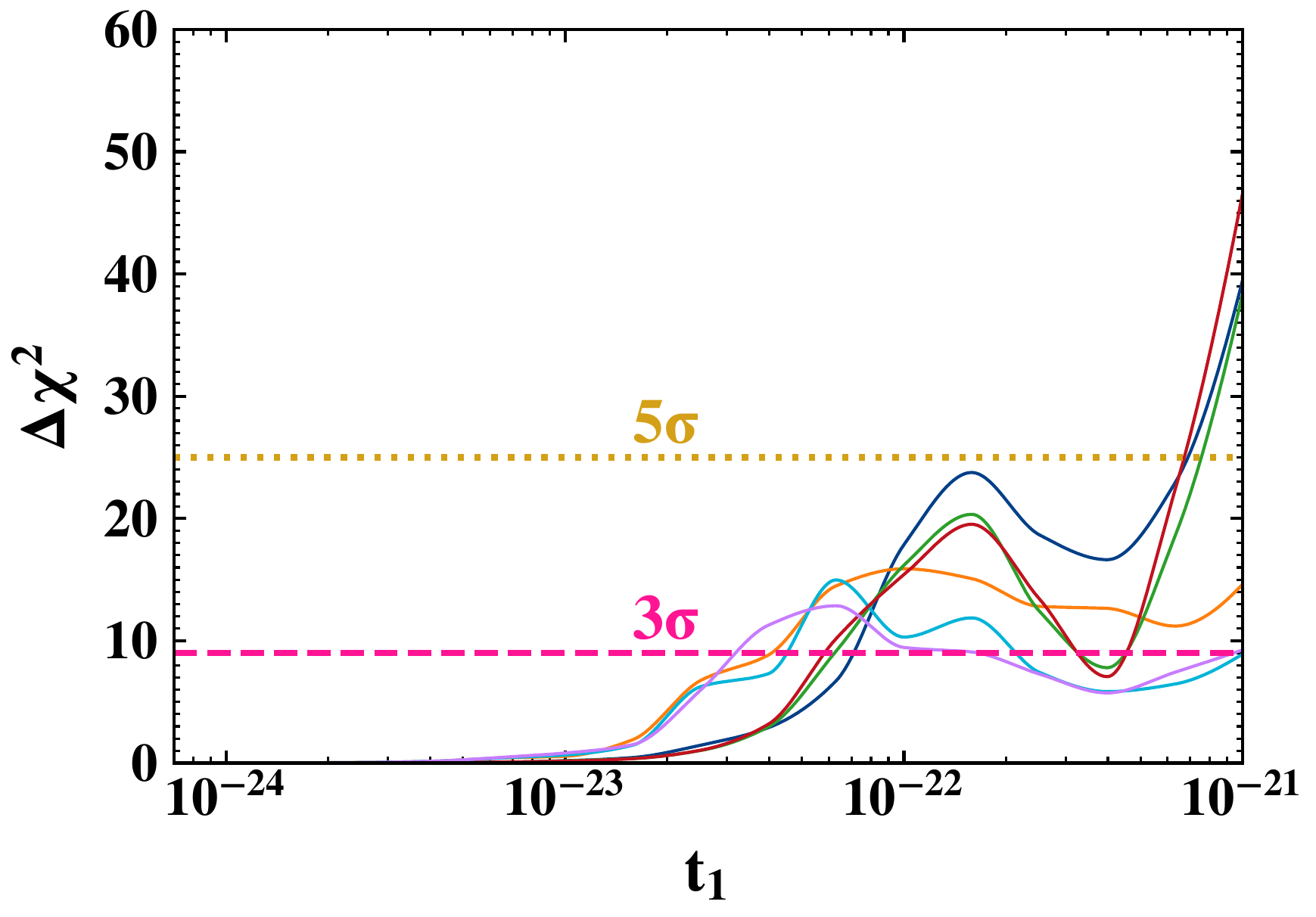}
     \includegraphics[width=0.49\linewidth]{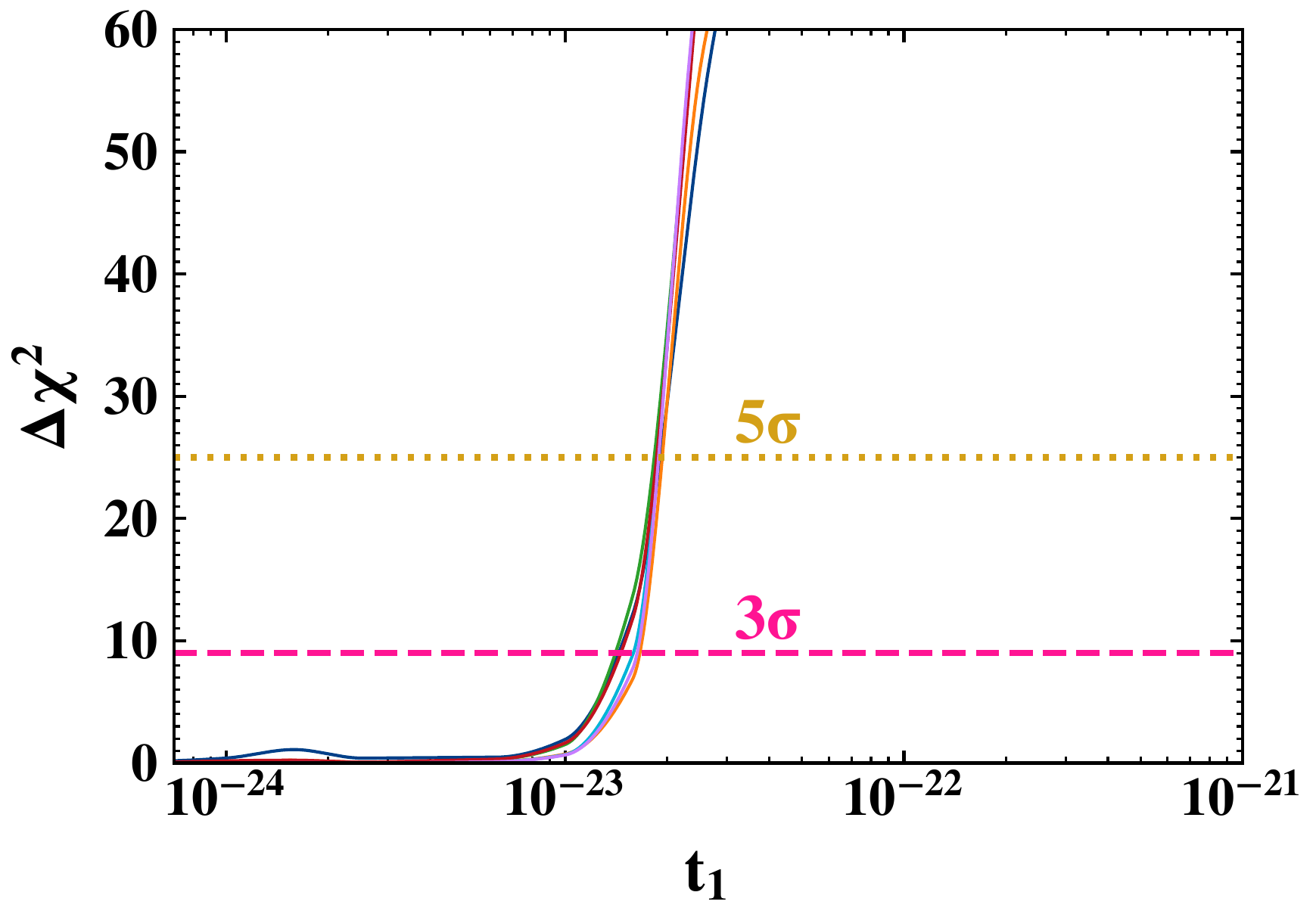}
 \vspace{0.3cm}

\begin{tikzpicture}

    \draw[line width=1.5pt, color=blue] (0,0) -- (0.7,0);
    \node[anchor=west] at (0.8,0)
    {$\mathrm{LE},\ \theta_{23}=43^\circ$};

    \draw[line width=1.5pt, color=green!60!black] (3.8,0) -- (4.5,0);
    \node[anchor=west] at (4.6,0)
    {$\mathrm{LE},\ \theta_{23}=45^\circ$};

    \draw[line width=1.5pt, color=red] (7.6,0) -- (8.3,0);
    \node[anchor=west] at (8.4,0)
    {$\mathrm{LE},\ \theta_{23}=47^\circ$};

    \draw[line width=1.5pt, color=orange] (0,-0.5) -- (0.7,-0.5);
    \node[anchor=west] at (0.8,-0.5)
    {$\mathrm{ME},\ \theta_{23}=43^\circ$};

    \draw[line width=1.5pt, color=cyan] (3.8,-0.5) -- (4.5,-0.5);
    \node[anchor=west] at (4.6,-0.5)
    {$\mathrm{ME},\ \theta_{23}=45^\circ$};

    \draw[line width=1.5pt, color=violet] (7.6,-0.5) -- (8.3,-0.5);
    \node[anchor=west] at (8.4,-0.5)
    {$\mathrm{ME},\ \theta_{23}=47^\circ$};

\end{tikzpicture}

\caption{Expected 1D $\Delta\chi^2$ exclusion plots as a function of the VEP parameter $t_1$, varied over the range $10^{-24}$--$10^{-21}$, contrasting the isolated appearance (left) and disappearance (right) channels. The performance of the Low-Energy (LE) flux configuration is compared with that of the Medium-Energy (ME) configuration for three true values of the atmospheric mixing angle: $\theta_{23} = 43^\circ$, $\theta_{23} = 45^\circ$, and $\theta_{23} = 47^\circ$. The other VEP parameter is fixed to zero ($t_2 = 0$).}
    \label{fig:app_disapp_1d}
\end{figure}

\subsubsection{One-Dimensional Sensitivity: Appearance vs. Disappearance Channels}
Fig~\ref{fig:app_disapp_1d}, we present the exclusion plots obtained by considering appearance (left) and disappearance (right) channels separately. The $\Delta\chi^2$ plots for the appearance channel rise gradually around $t_1\sim 10^{-23}$--$10^{-22}$, but the results show non-monotonic behaviour. The curves develop peaks and dips, indicating that the
exclusion sensitivity does not improve monotonically with increasing
$t_1$. The curves also depend noticeably on the beam configuration
(LE versus ME) and the true value of $\theta_{23}$. In contrast, the
disappearance channel curves remain close to zero over much of the
lower-$t_1$ range, followed by a steep rise around $t_1\sim 10^{-23}$.
This rise is nearly monotonic, with all six configurations following
almost the similar trajectory.

\begin{figure}
    \centering

    \includegraphics[width=0.49\linewidth]{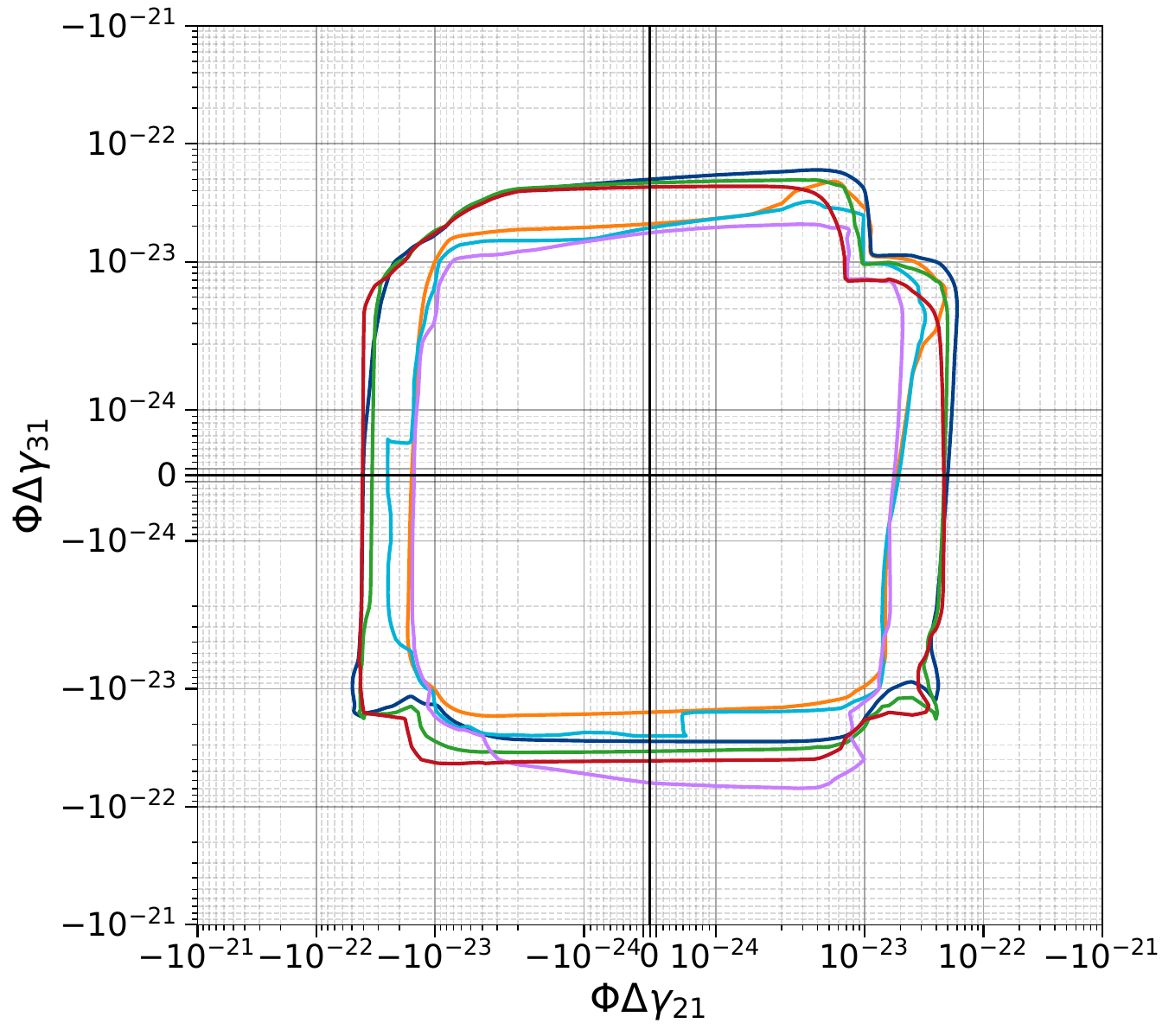}
     \includegraphics[width=0.49\linewidth]{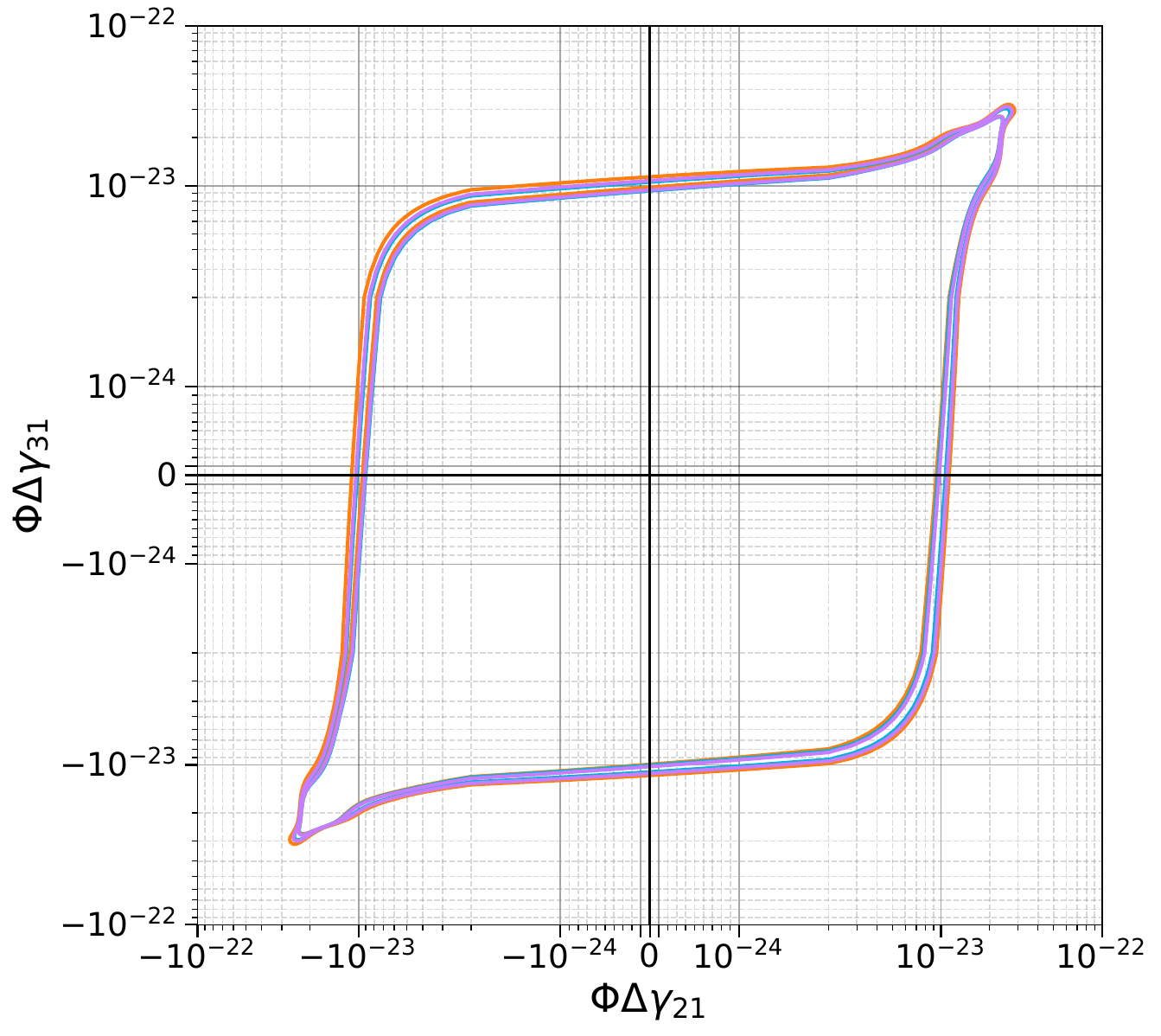}
 
\begin{tikzpicture}

    \draw[line width=1.5pt, color=blue] (0,0) -- (0.7,0);
    \node[anchor=west] at (0.8,0)
    {$\mathrm{LE},\ \theta_{23}=43^\circ$};

    \draw[line width=1.5pt, color=green!60!black] (3.8,0) -- (4.5,0);
    \node[anchor=west] at (4.6,0)
    {$\mathrm{LE},\ \theta_{23}=45^\circ$};

    \draw[line width=1.5pt, color=red] (7.6,0) -- (8.3,0);
    \node[anchor=west] at (8.4,0)
    {$\mathrm{LE},\ \theta_{23}=47^\circ$};

    \draw[line width=1.5pt, color=orange] (0,-0.5) -- (0.7,-0.5);
    \node[anchor=west] at (0.8,-0.5)
    {$\mathrm{ME},\ \theta_{23}=43^\circ$};

    \draw[line width=1.5pt, color=cyan] (3.8,-0.5) -- (4.5,-0.5);
    \node[anchor=west] at (4.6,-0.5)
    {$\mathrm{ME},\ \theta_{23}=45^\circ$};

    \draw[line width=1.5pt, color=violet] (7.6,-0.5) -- (8.3,-0.5);
    \node[anchor=west] at (8.4,-0.5)
    {$\mathrm{ME},\ \theta_{23}=47^\circ$};

\end{tikzpicture}

\caption{Expected 90\% C.L. sensitivity contours in the $\Phi\Delta\gamma_{21}$--$\Phi\Delta\gamma_{31}$ parameter space, contrasting the isolated appearance (left, varied over the range $-10^{-21}$ to $10^{-21}$) and disappearance (right, varied over the range $-10^{-22}$ to $10^{-22}$) channels. The performance of the Low-Energy (LE) flux configuration is compared with that of the Medium-Energy (ME) configuration for three true values of the atmospheric mixing angle: $\theta_{23} = 43^\circ$, $\theta_{23} = 45^\circ$, and $\theta_{23} = 47^\circ$. All standard oscillation parameters and unconstrained VEP parameters ($\theta_N, \Phi_N$) are marginalized.}
    \label{fig:app_disapp_2d}
      \end{figure}
      
\subsubsection{Two-Dimensional Contours: Appearance vs. Disappearance}
Fig~\ref{fig:app_disapp_2d} projects these isolated constraints onto the $\mathit{\Phi}\Delta\gamma_{21} - \mathit{\Phi}\Delta\gamma_{31}$ parameter space. The appearance-only allowed regions are visibly relaxed ($\sim 10^{-22}$) and exhibit irregular, jagged topologies due to unresolved interference effects involving $\delta_{\mathrm{CP}}$ and $\mathit{\Phi}_N$. In contrast, the disappearance-only contours drive the absolute statistical limits down to the $\sim 10^{-23}$ scale with much smoother boundaries, though they display a diagonal elongation indicating a correlated parameter dependence. Since the appearance channel is particularly sensitive to the standard CP-violating phase $\delta_{\mathrm{CP}}$, the interplay between $\delta_{\mathrm{CP}}$ and the additional VEP-related phase can introduce parameter degeneracies, thereby weakening the exclusion sensitivity and producing a non-monotonic $\Delta\chi^2$ region. In contrast, the disappearance channel exhibits a much weaker dependence on $\delta_{\mathrm{CP}}$, resulting in a comparatively smooth and nearly monotonic exclusion plots.

\subsection{Analysis based on Beam polarity}

Having established the complementary roles of the appearance and disappearance channels, we now partition the DUNE exposure by beam polarity to evaluate the distinct statistical contributions of the neutrino-only (FHC) and antineutrino-only (RHC) datasets.

\begin{figure}
    \centering

     \includegraphics[width=0.49\linewidth]{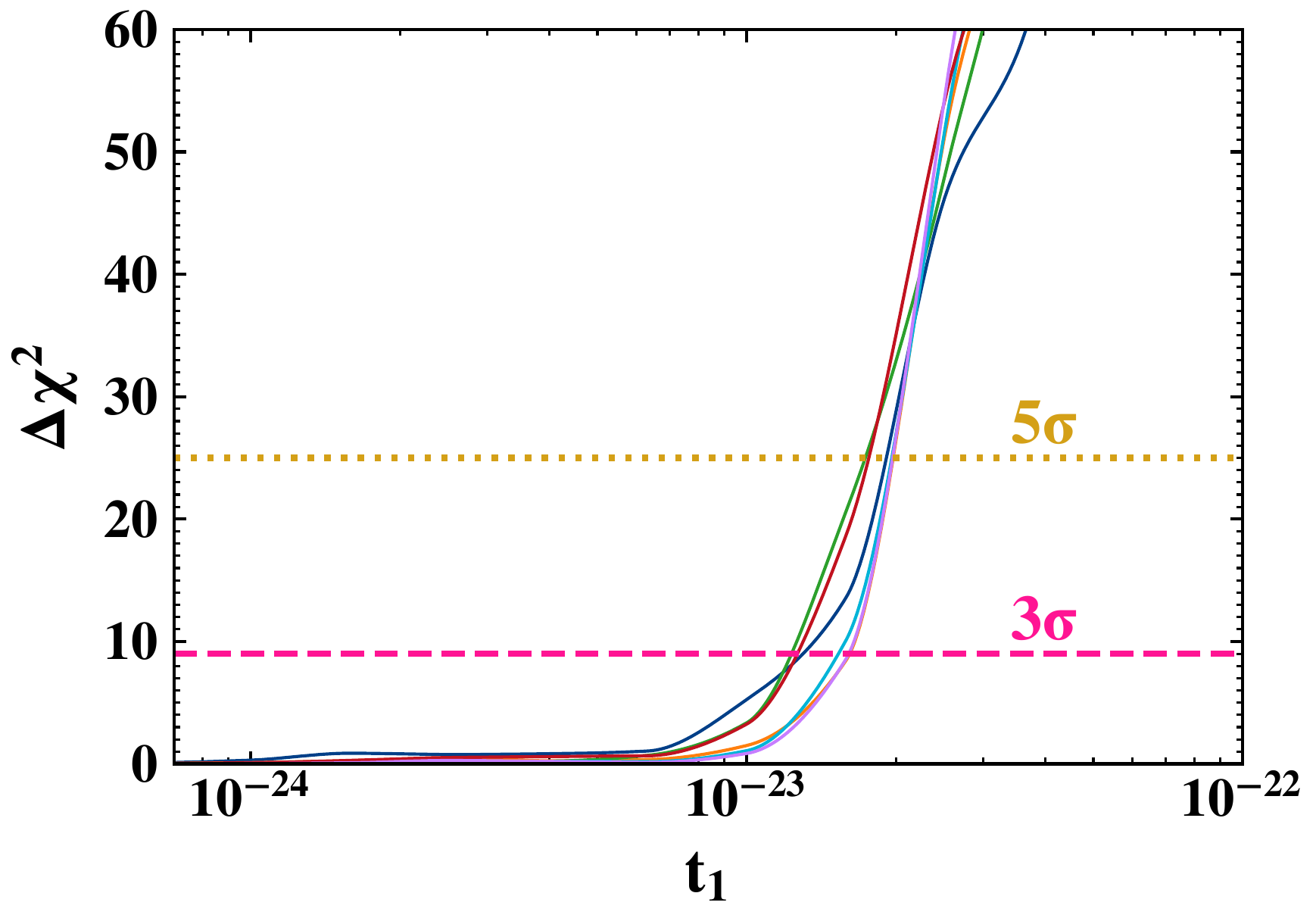}
     \includegraphics[width=0.49\linewidth]{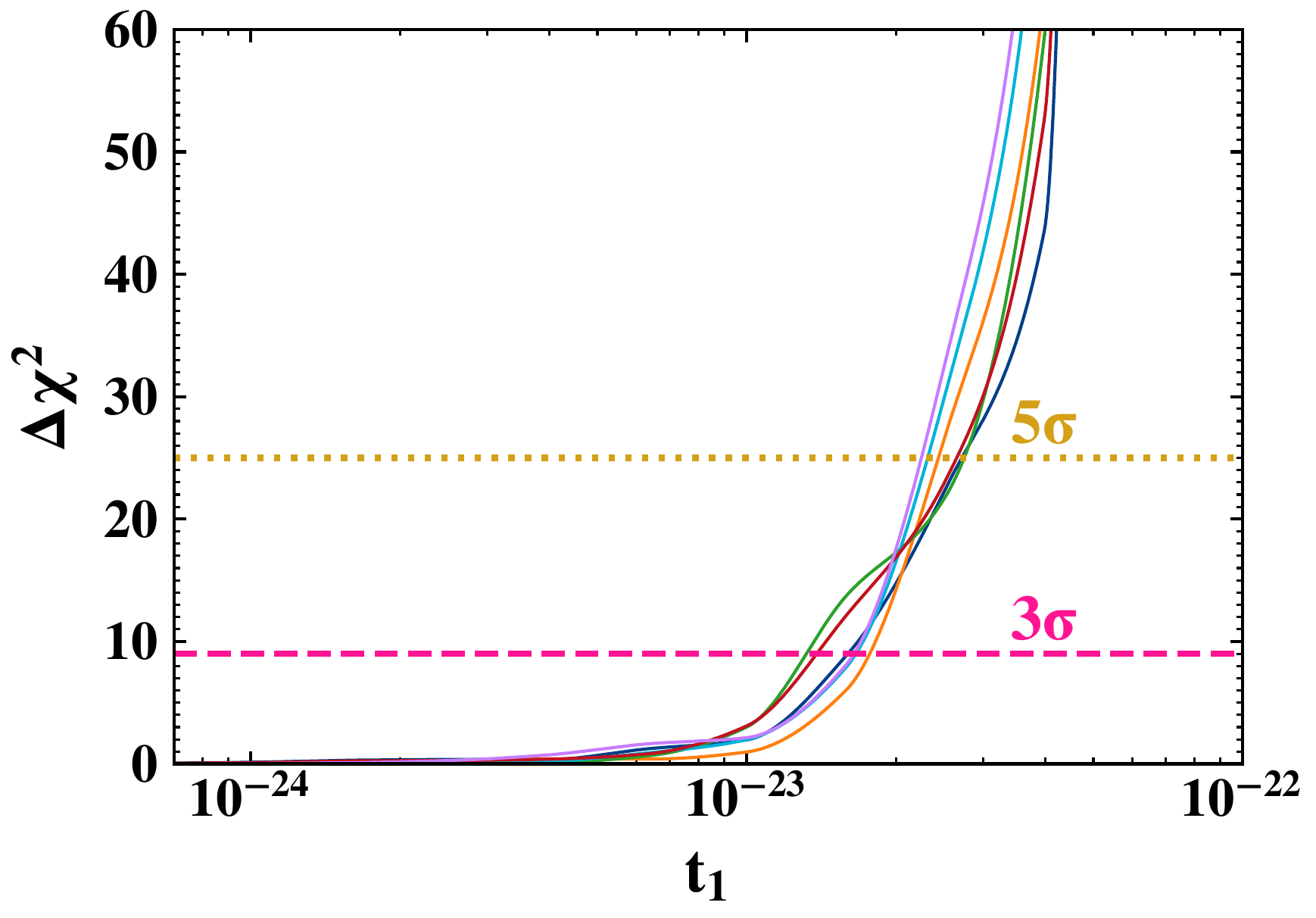}
            \vspace{0.3cm}

\begin{tikzpicture}

    \draw[line width=1.5pt, color=blue] (0,0) -- (0.7,0);
    \node[anchor=west] at (0.8,0)
    {$\mathrm{LE},\ \theta_{23}=43^\circ$};

    \draw[line width=1.5pt, color=green!60!black] (3.8,0) -- (4.5,0);
    \node[anchor=west] at (4.6,0)
    {$\mathrm{LE},\ \theta_{23}=45^\circ$};

    \draw[line width=1.5pt, color=red] (7.6,0) -- (8.3,0);
    \node[anchor=west] at (8.4,0)
    {$\mathrm{LE},\ \theta_{23}=47^\circ$};

    \draw[line width=1.5pt, color=orange] (0,-0.5) -- (0.7,-0.5);
    \node[anchor=west] at (0.8,-0.5)
    {$\mathrm{ME},\ \theta_{23}=43^\circ$};

    \draw[line width=1.5pt, color=cyan] (3.8,-0.5) -- (4.5,-0.5);
    \node[anchor=west] at (4.6,-0.5)
    {$\mathrm{ME},\ \theta_{23}=45^\circ$};

    \draw[line width=1.5pt, color=violet] (7.6,-0.5) -- (8.3,-0.5);
    \node[anchor=west] at (8.4,-0.5)
    {$\mathrm{ME},\ \theta_{23}=47^\circ$};

\end{tikzpicture}
\caption{Expected 1D $\Delta\chi^2$ exclusion plots as a function of the VEP parameter $t_1$, varied over the range $10^{-24}$--$10^{-22}$, contrasting the isolated neutrino-only (left) and antineutrino-only (right) operational modes. The performance of the Low-Energy (LE) flux configuration is compared with that of the Medium-Energy (ME) configuration for three true values of the atmospheric mixing angle: $\theta_{23} = 43^\circ$, $\theta_{23} = 45^\circ$, and $\theta_{23} = 47^\circ$. The other VEP parameter is fixed to zero ($t_2 = 0$), while all standard oscillation and exotic mixing parameters ($\theta_N, \Phi_N$) are marginalized.}
    
\label{fig:nu_anu_1d}

\end{figure}

\subsubsection{1D analysis}

 As in Fig.~\ref{fig:chi2_1d_full}, we fix $t_2=0$ and follow the same marginalization procedure over the remaining oscillation and VEP parameters to obtain the results presented in Fig.~\ref{fig:nu_anu_1d}. The left panel of Fig.~\ref{fig:nu_anu_1d} shows the exclusion plots the neutrino mode, while the right panel presents the corresponding results for the anti-neutrino mode. In neutrino mode, the LE configuration generally exhibits a stronger exclusion sensitivity, whereas in antineutrino mode, the broader ME spectrum provides enhanced sensitivity at the $5\sigma$ exclusion threshold. Notably, both running modes exhibit a steep rise in $\Delta\chi^2$. This demonstrates that DUNE can provide substantial constraints on the VEP parameters even when the full exposure is dedicated exclusively to neutrino or antineutrino running. The differences between the LE and ME configurations further illustrate the importance of the beam spectrum in determining the exclusion reach.

\begin{figure}
    \centering
    
     \includegraphics[width=0.49\linewidth]{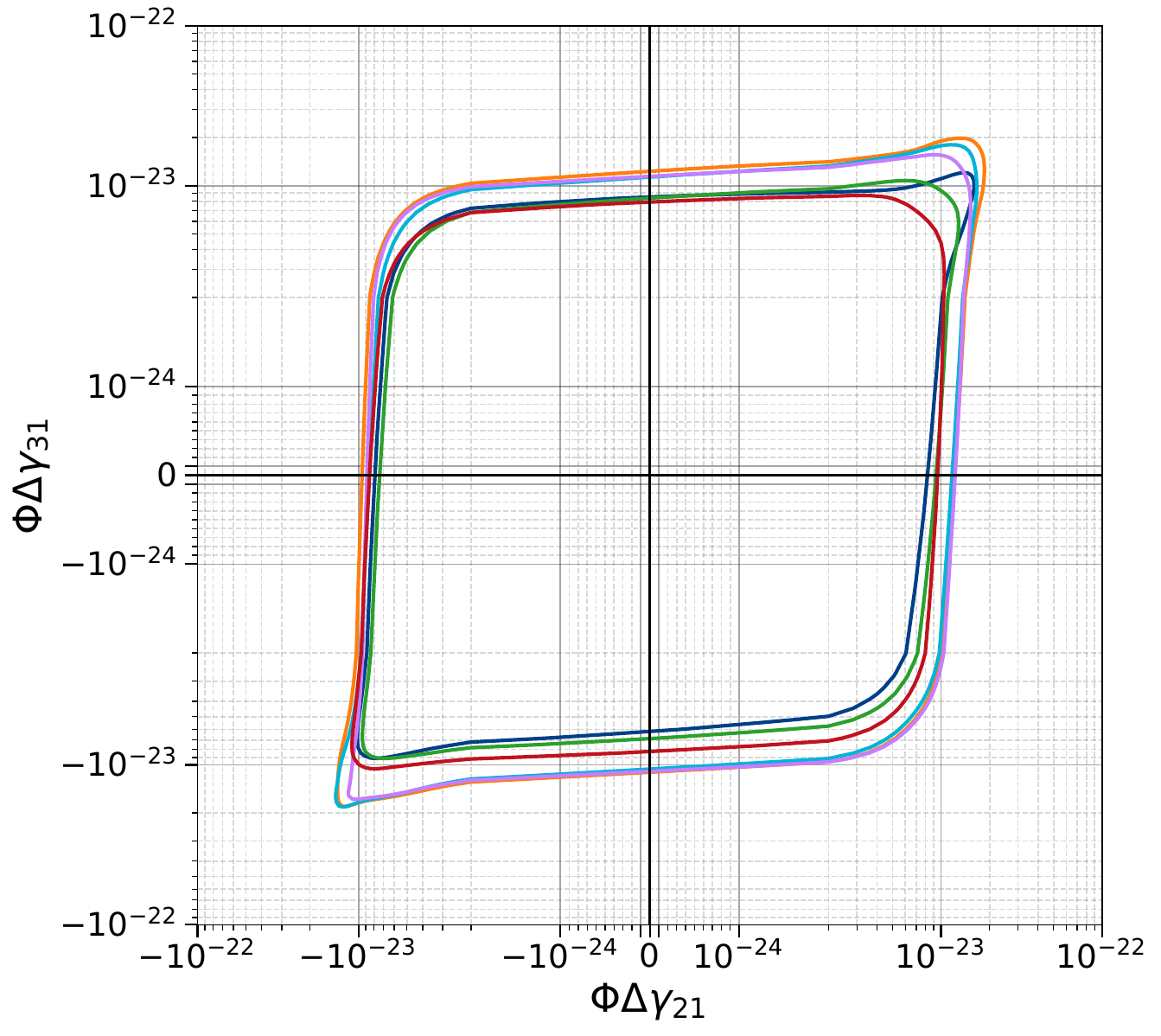}
     \includegraphics[width=0.49\linewidth]{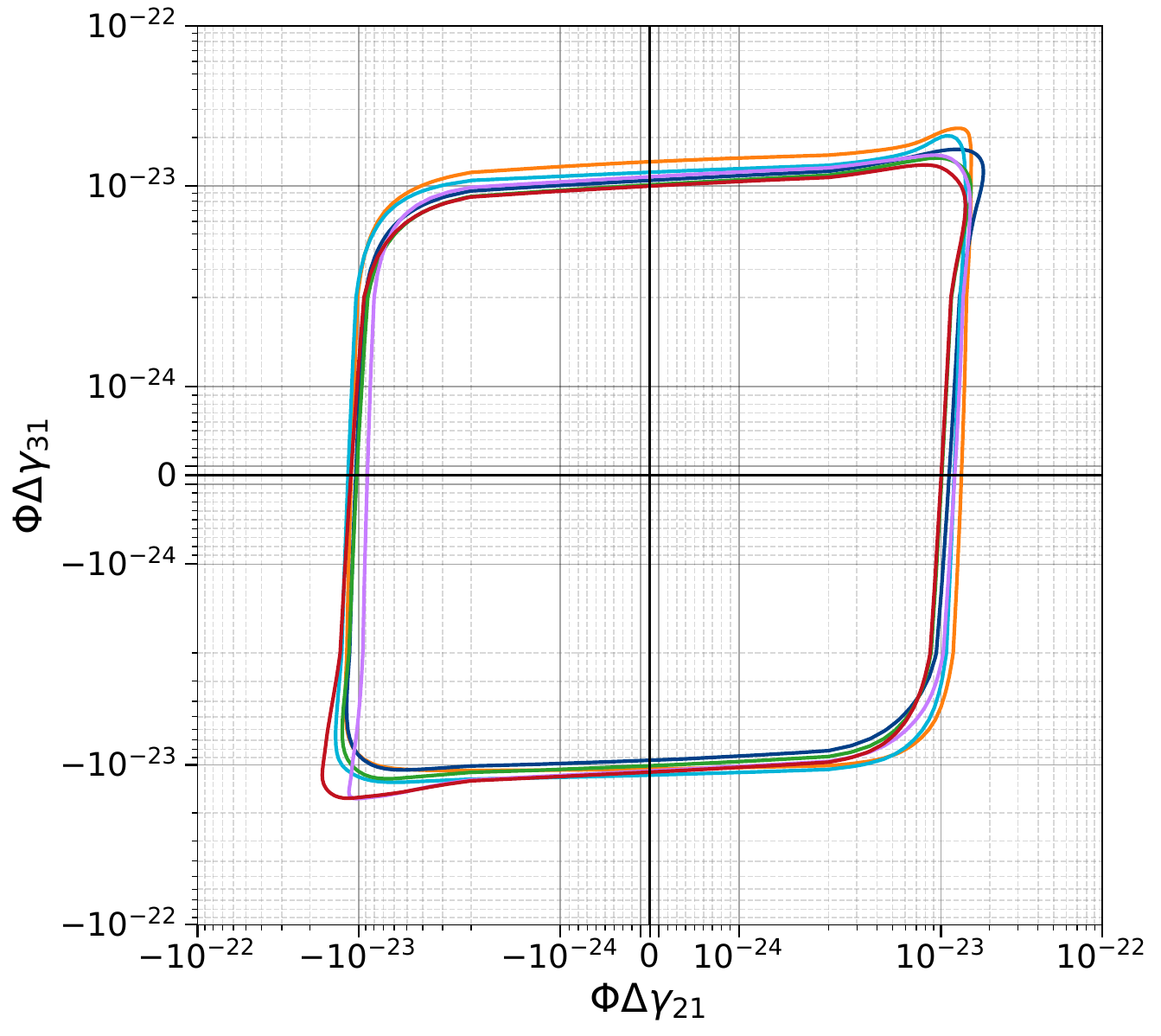}
        \vspace{0.3cm}

\begin{tikzpicture}

    \draw[line width=1.5pt, color=blue] (0,0) -- (0.7,0);
    \node[anchor=west] at (0.8,0)
    {$\mathrm{LE},\ \theta_{23}=43^\circ$};

    \draw[line width=1.5pt, color=green!60!black] (3.8,0) -- (4.5,0);
    \node[anchor=west] at (4.6,0)
    {$\mathrm{LE},\ \theta_{23}=45^\circ$};

    \draw[line width=1.5pt, color=red] (7.6,0) -- (8.3,0);
    \node[anchor=west] at (8.4,0)
    {$\mathrm{LE},\ \theta_{23}=47^\circ$};

    \draw[line width=1.5pt, color=orange] (0,-0.5) -- (0.7,-0.5);
    \node[anchor=west] at (0.8,-0.5)
    {$\mathrm{ME},\ \theta_{23}=43^\circ$};

    \draw[line width=1.5pt, color=cyan] (3.8,-0.5) -- (4.5,-0.5);
    \node[anchor=west] at (4.6,-0.5)
    {$\mathrm{ME},\ \theta_{23}=45^\circ$};

    \draw[line width=1.5pt, color=violet] (7.6,-0.5) -- (8.3,-0.5);
    \node[anchor=west] at (8.4,-0.5)
    {$\mathrm{ME},\ \theta_{23}=47^\circ$};

\end{tikzpicture}
\caption{Expected 90\% C.L. sensitivity contours in the $\Phi\Delta\gamma_{21}$--$\Phi\Delta\gamma_{31}$ parameter space, varied over the range $-10^{-22}$ to $10^{-22}$, contrasting the isolated neutrino-only (left) and antineutrino-only (right) operational modes. The performance of the Low-Energy (LE) flux configuration is compared with that of the Medium-Energy (ME) configuration for three true values of the atmospheric mixing angle: $\theta_{23} = 43^\circ$, $\theta_{23} = 45^\circ$, and $\theta_{23} = 47^\circ$. All standard oscillation parameters and unconstrained VEP parameters ($\theta_N, \Phi_N$) are marginalized.}
\label{fig:nu_anu_2d}
\end{figure}

\subsubsection{Two-Dimensional Contours: Neutrino vs. Antineutrino Modes}

Fig.~\ref{fig:nu_anu_2d} projects these isolated constraints onto the $\mathit{\Phi}\Delta\gamma_{21}-\mathit{\Phi}\Delta\gamma_{31}$ parameter space. The neutrino-only dataset independently breaks major phase-dependent degeneracies, restricting the allowed regions to smooth, rigid contours tightly bound around $\sim 10^{-23}$ scale. In contrast, the antineutrino mode provides weaker constraints, with broader allowed regions due to the reduced statistical power of the antineutrino mode. 

\section{ Summary and Discussion}
\label{sec:conclusion}

To contextualize the projected sensitivity of the DUNE experiment, Table \ref{tab:literature_comparison} summarizes the bounds obtained in this work alongside existing constraints established by previous atmospheric, solar, and accelerator experiments.

\begin{table*}[htbp]
\centering
\footnotesize
\caption{Comparison of the bounds obtained in this work with some of the already existing bounds from different atmospheric, solar, and accelerator experiments.}
\label{tab:literature_comparison}
\renewcommand{\arraystretch}{1.2} 
\begin{tabular}{lcc}
\hline\hline
Experiment / Reference & Parameter & Bounds \\
\hline
Atmospheric Data / Esmaili \textit{et al.} \cite{Esmaili:2014ota} & $\mathit{\Phi}\Delta\gamma_{21}$ & $\in [-9.2,\, 9.1] \times 10^{-27}$ (90\% C.L.) \\
Atmospheric Data / Esmaili \textit{et al.} \cite{Esmaili:2014ota} & $\mathit{\Phi}\Delta\gamma_{31}$ & $\in [-6.3,\, 5.6] \times 10^{-27}$ (90\% C.L.) \\
Atmospheric Data / Battistoni \textit{et al.} \cite{Battistoni2005} & $|\mathit{\Phi}\Delta\gamma_{32}|$ & $< 1.5 \times 10^{-25}$ (90\% C.L.) \\
Atmospheric Data / Gonzalez-Garcia \textit{et al.} \cite{Gonzalez-Garcia:2004pka} & $|\mathit{\Phi}\Delta\gamma_{32}|$ & $< 4.0 \times 10^{-25}$ ($3\sigma$) \\
Solar Data / Gago \textit{et al.} \cite{Gago:1999hi} & $|\mathit{\Phi}\Delta\gamma|$ & $\lesssim 10^{-13}$ (90\% C.L.) \\
Solar+KamLAND / Valdiviesso \textit{et al.} \cite{Valdiviesso2011} & $|\mathit{\Phi}\Delta\gamma_{21}|$ & $\lesssim 10^{-15}$ (90\% C.L.) \\
Accelerator Data / Mann \textit{et al.} \cite{Mann:1995nw} & $|\mathit{\Phi}\Delta\gamma_{e\mu}|$ & $< 2.3 \times 10^{-19}$ (90\% C.L.) \\
Accelerator Data / D\'{\i}az \textit{et al.} \cite{Diaz:2020aax} & $\mathit{\Phi}\Delta\gamma_{21}$ & $\in [-1.2,\, 1.4] \times 10^{-24}$ ($3\sigma$) \\
Accelerator Data / D\'{\i}az \textit{et al.} \cite{Diaz:2020aax} & $\mathit{\Phi}\Delta\gamma_{31}$ & $\in [-1.4,\, 1.2] \times 10^{-24}$ ($3\sigma$) \\
\hline
DUNE Sensitivity (This Work, LE $\theta_{23} = 43^\circ$) & $\mathit{\Phi}\Delta\gamma_{21}$ & $\in [-8.47,\, 8.47] \times 10^{-24}$ ($3\sigma$) \\
 & $\mathit{\Phi}\Delta\gamma_{31}$ & $\in [-8.53,\, 8.53] \times 10^{-24}$ ($3\sigma$) \\
 & $\mathit{\Phi}\Delta\gamma_{21}$ & $\in [-7.46,\, 9.53] \times 10^{-24}$ (90\% C.L.) \\
 & $\mathit{\Phi}\Delta\gamma_{31}$ & $\in [-5.86,\, 7.04] \times 10^{-24}$ (90\% C.L.) \\
\hline\hline
\end{tabular}
\end{table*}

In this work, we presented a comprehensive simulation-based evaluation of DUNE's expected exclusion sensitivity to a generalized VEP framework. Focusing on the exotic mass-coupling parameters $\mathit{\Phi}\Delta\gamma_{21}$ and $\mathit{\Phi}\Delta\gamma_{31}$, we systematically evaluated the performance of the Low-Energy (LE) and Medium-Energy (ME) beam configurations across a $13\text{--year}$ combined exposure ($6.5\,\text{yr FHC} + 6.5\,\text{yr RHC}$).

For the LE configuration, we find that, at $3\sigma$ confidence level, the projected allowed ranges for the assumed true value of $\theta_{23}=43^\circ$ are $\mathit{\Phi}\Delta\gamma_{21}\in[-8.47,\,8.47]\times10^{-24}$ and $\mathit{\Phi}\Delta\gamma_{31}\in[-8.53,\,8.53]\times10^{-24}$. These bounds are significantly tighter than those obtained for the ME configuration.

We also observe from this analysis that the appearance channels ($\nu_e, \bar{\nu}_e$), being phase-dependent, suffer from severe parameter degeneracies that produce irregular contour topologies as well as non-monotonic exclusion plots. In contrast, high-statistics disappearance channels ($\nu_\mu, \bar{\nu}_\mu$) act as a magnitude anchor, providing a monotonic $\Delta\chi^2$ rise that constrains parameter values down to the $10^{-23}$ level across all octants. 

The separate analyses of neutrino and antineutrino running further demonstrate that each mode can independently probe VEP effects at DUNE. Interestingly, in antineutrino mode, the ME configuration exhibits stronger exclusion sensitivity than the LE configuration at the expected $5\sigma$ threshold.

Analysis of the underlying oscillation probabilities and reconstructed event spectra confirms that VEP perturbations induce distinct energy-dependent spectral distortions centered near the first oscillation maximum. Because the LE flux concentrates its peak intensity directly within this kinematically sensitive $2\text{--}3\,\mathrm{GeV}$ window, it consistently yields tighter statistical bounds than the ME beam.

\section*{Acknowledgment}

DD acknowledges fruitful discussions with Prof. Raj Gandhi and Sandeep Sehrawat. Their insights and suggestions were helpful in carrying out this work.

\appendix
\section{Reconstructed Disappearance Event Spectra}
\label{app:disappearance}

\begin{figure}
    \centering
         \includegraphics[width=0.49\linewidth]{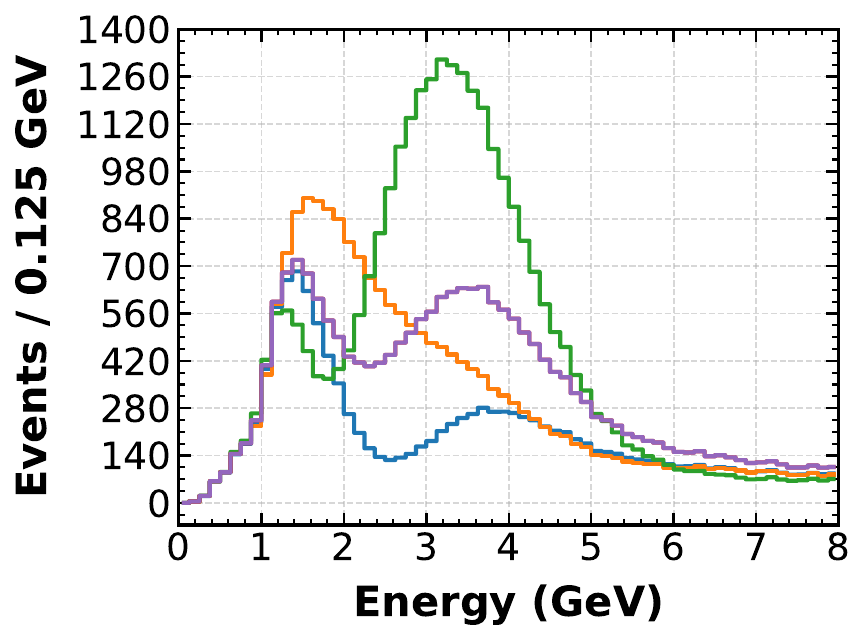}
    \includegraphics[width=0.49\linewidth]{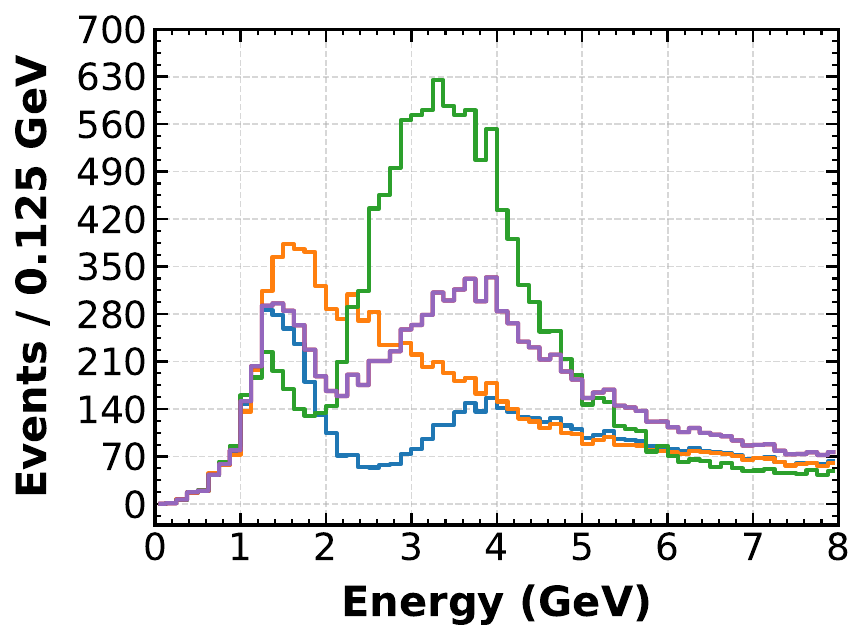}
    \includegraphics[width=0.49\linewidth]{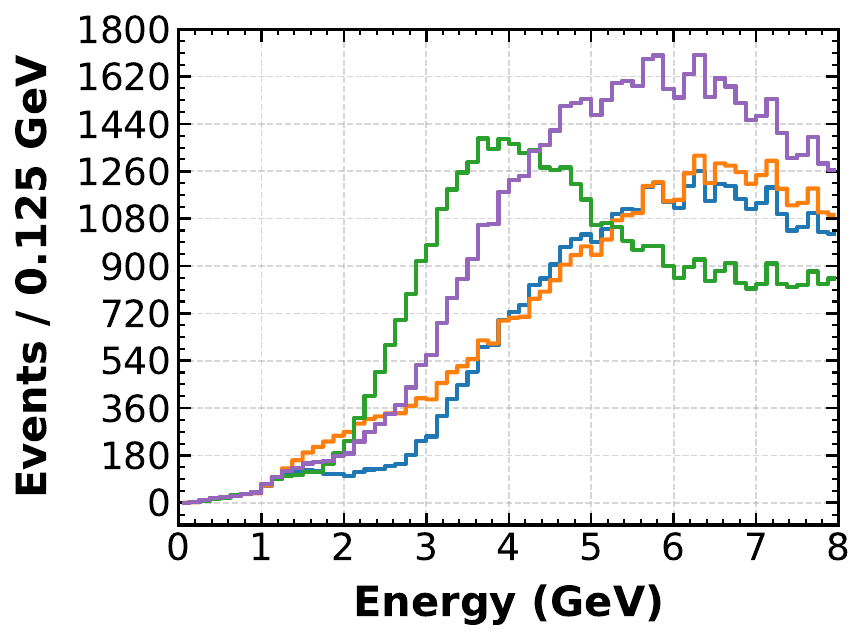}
    \includegraphics[width=0.49\linewidth]{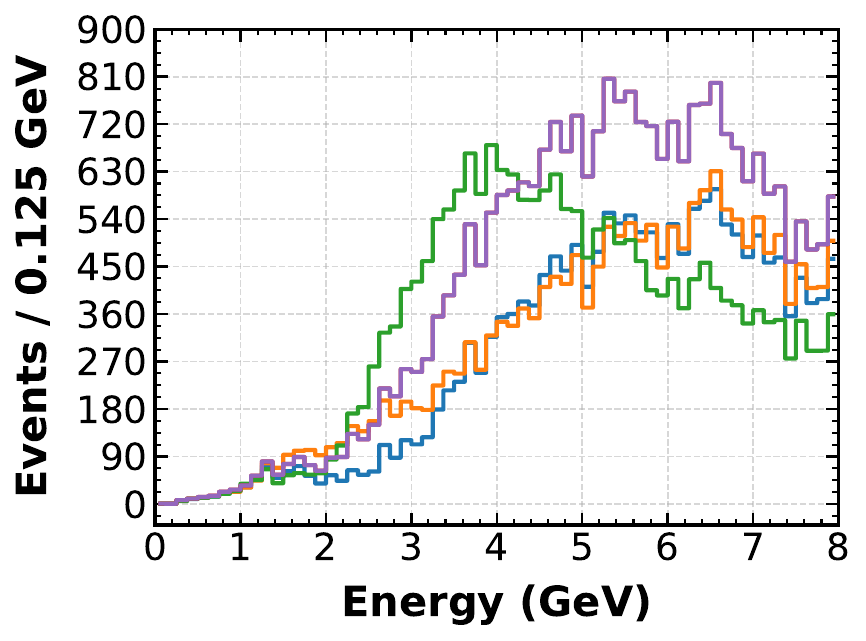}
      \includegraphics[width=0.60\textwidth]{legend.pdf}
\caption{Expected reconstructed disappearance event spectra at the $40\,\text{kton}$ DUNE Far Detector for a $6.5\,\text{yr}$ FHC + $6.5\,\text{yr}$ RHC exposure, comparing Low-Energy (top row) and Medium-Energy (bottom row) beam tunes. The top-left (top-right) panel shows the LE $\nu_\mu$ ($\bar{\nu}_\mu$) disappearance spectrum, while the bottom-left (bottom-right) panel shows the corresponding ME $\nu_\mu$ ($\bar{\nu}_\mu$) disappearance spectrum, all as functions of the reconstructed neutrino energy. }
          \label{fig:disappearance_events}
\end{figure}

To complement the appearance channel analysis, we evaluate the impact of VEP physics on the survival probabilities of muon neutrinos and antineutrinos. Reconstructed disappearance event spectra ($\nu_\mu \rightarrow \nu_\mu$ and $\bar{\nu}_\mu \rightarrow \bar{\nu}_\mu$) across the $40\,\mathrm{kton}$ DUNE Far Detector for a $6.5\,\text{yr FHC} + 6.5\,\text{yr RHC}$ exposure ($13\text{--year}$ total run) are shown in Fig.~\ref{fig:disappearance_events}.

Fig.~\ref{fig:disappearance_events} details the expected $\nu_\mu$ and $\bar{\nu}_\mu$ disappearance event spectra under both LE and ME beam tunes. The introduction of maximal exotic mixing ($\theta_N = 45^\circ$) drives significant high-energy event accumulation, particularly for $\mathit{\Phi}_N = 0^\circ$, which shifts the primary yield to a massive peak near $3.2\text{--}3.5\,\mathrm{GeV}$ in both neutrino and antineutrino modes. Conversely, the ME configuration resolves higher-energy disappearance features, where the standard spectra peak broadly between $6.0\text{--}6.5\,\mathrm{GeV}$. 

\bibliographystyle{unsrt}
\bibliography{references}

\end{document}